\documentclass[10pt, conference, compsocconf]{IEEEtran}
\IEEEoverridecommandlockouts 

\ifCLASSINFOpdf
\else
\fi
\usepackage[font=footnotesize]{subfig}

\usepackage{macros}
\usepackage{microtype}
\allowdisplaybreaks
\begin{document}
%
\title{A Note on the Complexity of the Satisfiability Problem for Graded Modal
  Logics}


\author{\IEEEauthorblockN{Yevgeny Kazakov}
  \IEEEauthorblockA{Computing Laboratory\\
    Oxford University\\
    Parks Rd., Oxford OX1 3QD, England\\
    e-mail: yevgeny.kazakov@comlab.ox.ac.uk} \and \IEEEauthorblockN{Ian
    Pratt-Hartmann} 
  \IEEEauthorblockA{
    School of Computer Science\\
    University of Manchester\\
    Oxford Rd., Manchester M13 9PL, England\\
    e-mail: ipratt@cs.man.ac.uk} }



\maketitle

\begin{abstract}
  \emph{Graded modal logic} is the formal language obtained from ordinary modal
  logic by endowing its modal operators with cardinality constraints. Under the
  familiar possible-worlds semantics, these augmented modal operators receive
  interpretations such as ``It is true at no fewer than 15 accessible worlds
  that \ldots'', or ``It is true at no more than 2 accessible worlds that
  \ldots''.  We investigate the complexity of satisfiability for this language
  over some familiar classes of frames.  This problem is more challenging than
  its ordinary modal logic counterpart---especially in the case of transitive
  frames, where graded modal logic lacks the tree-model property. We obtain
  tight complexity bounds for the problem of determining the satisfiability of a
  given graded modal logic formula over the classes of frames characterized by
  any combination of reflexivity, seriality, symmetry, transitivity and the
  Euclidean property.
\end{abstract}


%

\section{Introduction}
\label{sec:intro}

\emph{Graded modal logic} is the formal language obtained by decorating the
$\Diamond$-operator of ordinary modal logic with subscripts expressing
cardinality constraints. Specifically, for $C \geq 0$, the formula
$\Diamond_{\leq C} \phi$ may be glossed: ``$\phi$ is true at no more than $C$
accessible worlds,'' and the formula $\Diamond_{\geq C} \phi$ may be glossed:
``$\phi$ is true at no fewer than $C$ accessible worlds.''  The semantics for
graded modal logic generalize the relational semantics for ordinary modal logic
in the expected way. We employ the labels $\mbox{Rfl}$, $\mbox{Ser}$,
$\mbox{Sym}$, $\mbox{Tr}$ and $\mbox{Eucl}$ to denote, respectively, the classes
of reflexive, serial, symmetric, transitive and Euclidean frames. (Definitions
of these frame classes are given in Table~\ref{table:frameClasses}.) Using this
notation, $\bigcap \{ \mbox{Rfl}, \mbox{Tr}\}$ denotes the class of reflexive,
transitive frames, $\bigcap \{\mbox{Ser}, \mbox{Tr}, \mbox{Eucl}\}$ denotes the
class of serial, transitive, Euclidean frames, and so on. As a limiting case,
$\bigcap \emptyset$ denotes the class of all frames.  In this paper, we
investigate the computational complexity of determining the satisfiability of a
given formula of graded modal logic over any frame class of the form $\bigcap
\cF$, where $\cF \subseteq \{\mbox{Rfl}, \mbox{Ser}, \mbox{Sym}, \mbox{Tr},
\mbox{Eucl}\}$.

It is easy to see that ordinary modal logic is in effect a
sub-language of graded modal logic: any formula of the form $\Diamond
\phi$ may be equivalently written $\Diamond_{\geq 1} \phi$, and
similarly, any formula of the form $\Box \phi$ may be equivalently
written $\Diamond_{\leq 0} \neg \phi$.  And ordinary modal logic
provides a good starting point for our analysis, because its
complexity-theoretic treatment is comparatively straightforward. The
following two theorems are well-known, and may be proved using
techniques found in any modern text on modal logic
(e.g.~\cite{logic:BdRV}).  We remind the reader that symmetry and
transitivity together imply the Euclidean property.

\begin{theorem}\label{theo:simple}
  Let $\cF \subseteq \{\mbox{\rm Rfl}, \mbox{\rm Ser}, \mbox{\rm Sym}, \mbox{\rm
    Tr} , \mbox{\rm Eucl} \}$, with $\mbox{\rm Eucl} \in \cF$ or $\{\mbox{\rm
    Sym}, \mbox{\rm Tr}\} \subseteq \cF$.  Then the satisfiability problem for
  ordinary modal logic over $\bigcap \cF$ is $\NP$-complete.
\end{theorem}

\begin{theorem}\label{theo:ladner}
  If $\cF \subseteq \{\mbox{\rm Rfl}, \mbox{\rm Ser}, \mbox{\rm Tr} \}$, then
  the satisfiability problem for ordinary modal logic over $\bigcap \cF$ is
  $\PSPACE$-complete \textup{\cite{logic:ladner77}}. Also, if $\cF \subseteq
  \{\mbox{\rm Rfl}, \mbox{\rm Ser}, \mbox{\rm Sym} \}$, then the satisfiability
  problem for ordinary modal logic over $\bigcap \cF$ is $\PSPACE$-complete.
\end{theorem}

The upper complexity bound in Theorem~\ref{theo:simple} follows from the fact
that ordinary modal logic has the polynomial-size model property over the
relevant frame classes: if a formula $\phi$ of ordinary modal logic is
satisfiable over a frame in $\bigcap \cF$, where $\cF$ satisfies the conditions
of Theorem~\ref{theo:simple}, then it is satisfiable over a frame in $\bigcap
\cF$ whose size is bounded by a polynomial function of the number of symbols in
$\phi$.  For the frame classes of Theorem~\ref{theo:ladner}, ordinary modal
logic lacks the polynomial-size model property. However, it does have the
\emph{tree-model} property: if a formula is satisfiable over a frame in any of
the classes $\bigcap \cF$ mentioned in Theorem~\ref{theo:ladner}, then it is
satisfiable over a frame in that class which forms a (possibly infinite)
tree~\cite{logic:vardi97}.  Because the branches of this tree can be assumed to
be either short or periodic with small period, and because these branches can be
explored one-by-one, the $\PSPACE$-upper complexity bound may be obtained by
exhibiting, for each relevant frame class $\bigcap \cF$, a suitable semantic
tableau algorithm.

Turning our attention to the language of graded modal logic, our first
question is whether the results of Theorems~\ref{theo:simple}
and~\ref{theo:ladner} carry over to the larger language.  When $\cF$
contains neither of the classes $\mbox{Tr}$ or $\mbox{Eucl}$, the
answer is yes.  We have:
\begin{theorem}\label{theo:5:GMarbitrary}
  The satisfiability problem for graded modal logic over $\cF=\bigcap\emptyset$
  is $\PSPACE$-complete \textup{\cite{logic:tobies01}}.  In fact, if $\cF
  \subseteq \{\mbox{\rm Rfl}, \mbox{\rm Ser}, \mbox{\rm Sym} \}$, then the
  satisfiability problem for graded modal logic over $\bigcap \cF$ is
  $\PSPACE$-complete.
\end{theorem}
The reason---and indeed the reasoning---is essentially the same as for
Theorem~\ref{theo:ladner}: the $\PSPACE$ upper complexity bound in
Theorem~\ref{theo:5:GMarbitrary} depends on the fact that graded modal
logic enjoys the tree-model property over the relevant frame
classes. This can then be used to establish the correctness of
semantic tableau algorithms for graded modal logic over these frame
classes. The paper~\cite{logic:tobies01} actually considers only the
case $\cF = \emptyset$ (i.e. the class of all frames); however, the
modifications required to take account of reflexivity, seriality and
symmetry are routine, because these
restrictions do not compromise the tree-model property. Note that the
upper complexity bound in Theorem~\ref{theo:5:GMarbitrary} holds even
when numerical subscripts are coded in binary. (The much easier result
for unary coding can be found in~\cite{HollunderB91}.)

When $\cF$ contains either $\mbox{Eucl}$ or $\mbox{Tr}$, the complexity of the
satisfiability problem for graded modal logic over $\bigcap \cF$ is harder to
determine. Consider first the analogue of Theorem~\ref{theo:simple}, where we
have either $\mbox{Eucl} \in \cF$ or $\{\mbox{Tr}, \mbox{Sym}\} \subseteq \cF$,
and let $\{\phi_n\}_{n \geq 0}$ be the sequence of formulas given by $\phi_n =
\Diamond_{\geq 2^n} p$. Assuming binary coding of numerical subscripts, the
number of symbols in $\phi_n$ is bounded by a linear function of $n$, and every
$\phi_n$ is satisfiable over a Euclidean frame; but $\phi_n$ is certainly not
satisfiable over any frame with fewer than $2^n$ worlds! Thus, for graded modal
logic, the reasoning used to prove Theorem~\ref{theo:simple} fails.
Nevertheless, the corresponding complexity result still holds:
\begin{theorem}\label{theo:main1}
Let $\cF \subseteq \{\mbox{\rm Rfl}, \mbox{\rm Ser}, \mbox{\rm Sym},
\mbox{\rm Tr}, \mbox{\rm Eucl} \}$, with $\mbox{\rm Eucl} \in \cF$ or
$\{\mbox{\rm Sym}, \mbox{\rm Tr}\} \subseteq \cF$. Then the
satisfiability problem for graded modal logic over $\bigcap \cF$ is
$\NP$-complete.
\end{theorem}
\noindent
We prove Theorem~\ref{theo:main1} in Section~\ref{sec:euclidean}.

When $\cF$ contains $\mbox{Tr}$, but neither $\mbox{Sym}$ nor $\mbox{Eucl}$, we
cannot apply the reasoning of Theorem~\ref{theo:ladner} at all, since graded
modal logic lacks the tree-model property over transitive frames. For example,
consider the formula $\phi$ given by
\begin{equation*}
  \phi := q_0  \wedge 
  \Diamond_{\geq 2} (\neg q_0 \wedge q_1 \wedge
  \Diamond_{\geq 1} (\neg q_0 \wedge \neg q_1))\wedge 
  \Diamond_{\leq 1} \neg q_1.
\end{equation*}

The formula $\phi$ is certainly satisfiable over transitive frames; however, it
is not satisfiable over tree-shaped transitive frames.  For suppose $\phi$ is
true at a world $w_0$ in some structure.  The conjunct $\Diamond_{\geq 2} (\neg
q_0 \wedge q_1 \wedge \Diamond_{\geq 1} (\neg q_0 \wedge \neg q_1))$ ensures the
existence of distinct worlds $w_1$ and $w_2$, accessible from (and distinct
from) $w_0$, and, for $i = 1,2$, a world $w'_i$ accessible from $w_i$ and
satisfying $\neg q_1$, with $w'_i$ distinct from $w_0$, $w_1$ and $w_2$.  But
the conjunct $\Diamond_{\leq 1} \neg q_1$ ensures that, if the accessibility
relation is transitive, $w'_1 = w'_2$. Hence, $\phi$ is not satisfiable over a
tree.  Indeed, we show below that, for the relevant frame classes, graded modal
logic and ordinary modal logic exhibit different complexities:
\begin{theorem}\label{theo:main2}
  Let $\cF \subseteq \{\mbox{\rm Rfl}, \mbox{\rm Ser}, \mbox{\rm Tr} \}$, with
  $\mbox{\rm Tr} \in \cF$. Then the satisfiability problem for graded modal
  logic over $\bigcap \cF$ is $\NEXPTIME$-complete.  It remains
  $\NEXPTIME$-hard, even when all numerical subscripts in modal operators are
  at most $1$.
\end{theorem}
\noindent
We prove Theorem~\ref{theo:main2} in Section~\ref{sec:transitive}.  The final
statement of the theorem is significant, because it means that the result does
not depend upon the coding of numerical subscripts.

A moment's thought shows that the conditions in
Theorems~\ref{theo:5:GMarbitrary}--\ref{theo:main2} are exhaustive: together,
they establish the complexity of the satisfiability problem for graded modal
logic over $\bigcap \cF$ for every $\cF \subseteq \{\mbox{Rfl}, \mbox{Ser},
\mbox{Sym}, \mbox{Tr},\mbox{Eucl}\}$. 

The {\em decidability} of the satisfiability problem for graded modal logic over
various frame classes $\bigcap \cF$ is touched on in~\cite{logic:fine72}, where
it is stated (p.~520) that ``standard techniques or modifications of them may be
used to prove the decidability of most of [these] logics''; however, the paper
gives no further details.  Several such decidability results are claimed
in~\cite{logic:GMLV}; however, in the (difficult) case where $\cF =
\{\mbox{Tr}\}$, this proof contains an error, as reported
in~\cite{logic:ksz07}. The latter provides a correct proof; however, the method
employed there does not establish any complexity bounds. It is conjectured
in~\cite{HoekR95} (Remark~4.12), that the satisfiability problem for graded
modal logic over the class of transitive, symmetric and reflexive frames is
$\PSPACE$-complete: Theorem~\ref{theo:main1} shows that this conjecture, if
true, would imply that $\PSPACE$=$\NP$.  Earlier accounts of graded modal logics
focused primarily on the problem of axiomatizing the set of valid formulas over
these frame classes.
For instance, \cite{logic:fine72} provides (or reports) such
axiomatizations for $\bigcap \cF$, where $\cF$ is any of $\emptyset$,
$\{\mbox{Rfl}\}$, $\{\mbox{Sym}\}$, $\{\mbox{Rfl}, \mbox{Sym}\}$, $\{\mbox{Rfl},
\mbox{Tr}\}$ and $\{\mbox{Rfl}, \mbox{Tr}, \mbox{Sym}\}$. Similar results can be
found in~\cite{logic:GMLI,logic:GMLII,logic:GMLIII,logic:GMLIV}; see
also~\cite{HoekR95} for axiomatizations of some related logics.

Graded modal logics are closely related to terminological languages
and description logics (DLs)~\cite{dlhb2} featuring so-called
qualified number restrictions. These logics allow concepts to be
defined by specifying how many things (of various kinds) instances of
those concepts can be related to.  
Logics featuring both qualified number restrictions and transitive relations are
frequently undecidable~\cite{logic:Practical}, and many DLs incorporate various
syntactic restrictions to restore decidability.  It was recently shown
in~\cite{logic:ksz07} that some of these syntactic restrictions can be
considerably relaxed.

This paper is an extended version of \cite{logic:kp-h09:lics} containing the
omitted proofs.

\section{Preliminaries}
\label{sec:preliminaries}

Fix a countably infinite set $\Pi$. The language of \emph{graded modal
  logic} is defined to be the smallest set of expressions, $\cGM$,
satisfying the following conditions:
\begin{enumerate}
\item $\Pi \subseteq \cGM$;
\item if $\phi$ and $\psi$ are in $\cGM$, 
then so are $\neg \phi$, $\phi \wedge \psi$, $\phi
\vee \psi$, $\phi \rightarrow \psi$ and $\phi \leftrightarrow \psi$;
\item if $\phi$ is in $\cGM$, then so are $\Diamond_{\leq C} \phi$ and
  $\Diamond_{\geq C} \phi$, for any bit-string $C$.
\end{enumerate}
We refer to expressions in this set as $\cGM$-\emph{formulas} (or simply
\emph{formulas}, if clear from context).  If $\phi$ is a $\cGM$-formula, we take
the \emph{size of} $\phi$, denoted $\sizeof{\phi}$, to be the number of symbols
in $\phi$.  Throughout the paper, we equivocate between bit-strings and the
natural numbers they represent in the usual way.  Thus, we may informally think
of the subscripts in $\Diamond_{\leq C}$ and $\Diamond_{\geq C}$ as natural
numbers, it being understood that the number of symbols in, for example,
$\Diamond_{\leq C}$ is approximately $\log C$, rather than $C$. That is: in
giving the size of a formula, we assume \emph{binary}, rather than {\em unary},
coding.

Let $\Sigma$ be the relational signature with unary predicates $\Pi$
and single binary predicate $r$, and let $\fA$ be a $\Sigma$-structure
with domain $W$. We refer to the elements of $W$ as \emph{worlds}.
We
define the \emph{satisfaction} relation for $\cGM$-formulas
inductively as follows:
\begin{enumerate}
\item $\fA \models_w p$ if and only if $w \in p^\fA$;
\item $\fA \models_w \neg \phi$ if and only if $\fA \not \models_w
  \phi$, and similarly for $\wedge$, $\vee$, $\rightarrow$,
  $\leftrightarrow$;
\item $\fA \models_w \Diamond_{\geq C} \phi$ if and only if there
  exist at least $C$ worlds $v \in W$ such that $\langle w,v \rangle
  \in r^\fA$ and $\fA \models_v \phi$;
\item $\fA \models_w \Diamond_{\leq C} \phi$ if and only if there
  exist at most $C$ worlds $v \in W$ such that $\langle w,v \rangle
  \in r^\fA$ and $\fA \models_v \phi$.
\end{enumerate}
The notion of satisfaction extends to sets of $\cGM$-formulas $\Phi$
as expected: $\fA \models_w \Phi$ if $\fA \models_w \phi$ for all
$\phi \in \Phi$.  If $\fA \models_w \phi$, we sometimes say,
informally, that $\phi$ is \emph{true at} $w$ \emph{in} $\fA$.  We
write $\Box \phi$ as an abbreviation for $\Diamond_{\leq 0} \neg
\phi$, and $\Diamond \phi$ as an abbreviation for $\Diamond_{\geq 1}
\phi$, or, equivalently, $\neg \Diamond_{\leq 0} \phi$. Thus, the
language of ordinary modal logic may be regarded as the subset of
$\cGM$ in which all indices are restricted to 0. Finally, we write
$\boxdot\phi$ as an abbreviation for $\phi \wedge \Box \phi$.

By a \emph{frame}, we mean an $\{r\}$-structure---in other words, a
non-empty (possibly infinite) digraph.  If $\fA$ is a
$\Sigma$-structure, then its $\{r\}$-reduct is a frame $\fF$: we say
that $\fA$ is a structure \emph{over} $\fF$. Further, we call the
mapping $V: \Pi \rightarrow \bP(W)$ given by $p \mapsto p^\fA$ the
\emph{valuation} of $\fA$ (\emph{on} $W$).  We write $\fA = (W,R,V)$
to indicate that $\fA$ is a $\Sigma$-structure over the frame $(W,R)$
with valuation $V$. Obviously, this determines $\fA$ completely.
Henceforth, the term ``structure'', with no signature qualification,
will always mean ``$\Sigma$-structure''.  Let $\phi$ be a
$\cGM$-formula.  We say that $\phi$ is \emph{satisfiable over} a frame
$\fF$ if there exists a structure $\fA$ over $\fF$ and a world $w$ of
$\fA$ such that $\fA \models_w \phi$.  Further, $\phi$ is
\emph{satisfiable over} a class of frames $\cK$ if it is satisfiable
over some frame in $\cK$. We denote by $\cGM_\cK$-\emph{Sat} the
problem of determining whether a given $\cGM$-formula is satisfiable
over $\cK$.

Any first-order sentence $\alpha$ over the signature $\{r\}$ defines a class of
frames $\{\fF: \fF \models \alpha\}$. The most common frame classes are those
which we agreed in Section~\ref{sec:intro} to denote by the labels $\mbox{Rfl}$,
$\mbox{Ser}$, $\mbox{Sym}$, $\mbox{Tr}$ and $\mbox{Eucl}$.
Table~\ref{table:frameClasses} lists these frame classes together with their
respective defining first-order sentences.
\begin{table}
\caption{Frame classes considered in this paper.}
\centering
{\small
\begin{tabular}{ll}
reflexive frames & $\forall x. r(x,x)$\\
serial frames & $\forall x\exists y. r(x,y)$\\
symmetric frames & $\forall x \forall y. (r(x,y) \rightarrow r(y,x))$\\
transitive frames & $\forall x \forall y \forall z. (r(x,y) \wedge r(y,z)
                    \rightarrow r(x,z))$\\
Euclidean frames & $\forall x \forall y \forall z. (r(x,y) \wedge r(x,z)
                    \rightarrow r(y,z))$.
\end{tabular}
}
\label{table:frameClasses}
\end{table}
A structure over a reflexive frame will simply be called a {\em
  reflexive} structure, and similarly for the other frame properties.
We can now articulate the objective of this paper. Let $\cF$ be a
subset (possibly empty) of the set of frame classes $\{\mbox{Rfl},
\mbox{Ser}, \mbox{Sym}, \mbox{Tr}, \mbox{Eucl} \}$. We ask: what is
the complexity of $\cGM_{\cap\cF}$-Sat?

\section{Euclidean frames}
\label{sec:euclidean}

The purpose of this section is to prove Theorem~\ref{theo:main1}.  We make use
of a known complexity result on first-order logic with counting quantifiers.
Denote by $\cC^1$ the set of first-order formulas featuring only a single
variable $x$, but with the counting quantifiers $\exists_{\leq C} x$ and
$\exists_{\geq C} x$ allowed. The following result holds for both unary and
binary coding of numerical subscripts:
\begin{theorem}[\cite{logic:kuncak+rinard07,logic:ph08}]\label{theo:C1}
  The problem of deciding satisfiability for $\cC^1$-formulas is $\NP$-complete.
\end{theorem}
\noindent
We show that, for $\cGM$-formulas, satisfiability over Euclidean
frames is equivalent to satisfiability over frames having a
particularly simple form, and that, for such frames, the fragment
$\cC^1$ is as expressive as we need.

Let $\fF = (W,R)$ be a frame.  If $X \subseteq W$, $R(X)$
denotes $\bigcup_{x\in X}\set{w\in W\mid\tuple{x,w}\in
  R}$; we write $R(w)$ for $R(\{w\})$.  If $\fF = (W,R)$ is a frame,
and $X \subseteq W$, $R^*(X)$ denotes $X \cup R(X) \cup R(R(X)) \cup
\cdots$; we write $R^*(w)$ for $R^*(\{w\})$.  If $\fA$ is a structure
over a frame $(W,R)$ and $X \subseteq W$, let $\fB$ be the
substructure of $\fA$ with domain $R^*(X)$. We call $\fB$ the
\emph{substructure} generated by $X$. Note that reflexivity,
seriality, symmetry, transitivity and the Euclidean property are all
preserved under generated substructures.
\begin{lemma}\label{lma:generated}
Let $\phi$ be a formula of $\cGM$, $\fA$ a structure, $w$ a world of
$\fA$ and $\fB$ the substructure generated by $\{w\}$. If $\fA
\models_w \phi$, then $\fB \models_w \phi$.
\end{lemma}
\begin{IEEEproof}
Induction on the structure of $\phi$.
\end{IEEEproof}

\begin{lemma}\label{lma:totalOrCone}
  Let $\fF = (W,R)$ be a Euclidean frame and $w_0 \in W$. Then:
  \textup{(}i\textup{)} $R(w_0) \subseteq R(R(w_0))$, \textup{(}ii\textup{)}
  $R^*(w_0) = \{w_0\} \cup R(R(w_0))$, and \textup{(}iii\textup{)} $R$ is total
  on $R(R(w_0))$.
\end{lemma}

\begin{IEEEproof}
  For the first statement, observe that, in a Euclidean frame, $R$ is total on
  any set $R(w_0)$. In particular, $\langle w, w \rangle \in R$ for all $w \in
  R(w_0)$, whence $R(w_0) \subseteq R(R(w_0))$.

  Now consider any $X \subseteq W$ such that $R$ is total on $X$. We claim that
  $R$ is also total on $R(X)$, and that $R(X) = R(R(X))$.  By the Euclidean
  property, $\langle w,w \rangle \in R$ for all $w \in R(X)$, so that $R(X)
  \subseteq R(R(X))$. We show that $R$ is total on $R(X)$.  If $w \in R(X)$ and
  $R$ is total on $X$, then by the Euclidean property, $\langle x, w \rangle \in
  R$ for all $x \in X$, whence, if $w' \in R(X)$, using the Euclidean property
  again, $\langle w, w' \rangle \in R$. Thus $R$ is total on $R(X)$.  Finally,
  we show that $R(R(X)) \subseteq R(X)$.  Suppose $w \in R(R(X))$, so that
  $\langle w',w \rangle \in R$ for some $w' \in R(X)$.  Pick any $x \in X$.
  Since $R$ is total on $R(X) \supseteq X$, $\langle w',x \rangle \in R$, and
  so, by the Euclidean property, $\langle x,w \rangle \in R$. Thus, $R(R(X))
  \subseteq R(X)$, proving the claim.

  For the second statement of the lemma, putting $X = R(w_0)$ in the claim of
  the previous paragraph, we have $R(R(w_0)) = R(R(R(w_0))) = R(R(R(R(w_0)))) =
  \ldots$. Thus,
\begin{eqnarray*}
  R^*(w_0) &  = & \{w_0\} \cup R(w_0) \cup R(R(w_0)) \cup \cdots\\
  \  &  = & \{w_0\} \cup R(w_0) \cup R(R(w_0)) \\
  \  &  = & \{w_0\} \cup R(R(w_0)),
\end{eqnarray*}
with the last step following from the first statement of the lemma.
\end{IEEEproof}

Lemmas~\ref{lma:generated} and~\ref{lma:totalOrCone} show that, when discussing
satisfiability over Euclidean frames, we may restrict attention to frames of the
form $(W \cup \{w_0\},R)$, where $R$ is total on $W$, $R(w_0) \subseteq W$, and
$w_0$ may or may not be in $W$.  Over such simple frames, any $\cGM$-formula can
be translated into an equisatisfiable $\cC^1$-formula. Specifically:

\begin{lemma}\label{lma:c1:translation}
  Let $\cF \subseteq \{\mbox{\rm Rfl}, \mbox{\rm Ser}, \mbox{\rm Sym}, \mbox{\rm
    Tr} \}$.  Given a $\cGM$-formula $\phi$, we can compute, in time bounded by
  a polynomial function of $\sizeof{\phi}$, a $\cC^1$-formula $\alpha$ such that
  $\phi$ is satisfiable over a frame in $\bigcap \cF \cap \mbox{\rm Eucl}$ if
  and only if $\alpha$ is satisfiable.
\end{lemma}

\begin{IEEEproof}
Let $q_0$, $q_1$, $q_2$ be new unary predicates (i.e., pairwise
distinct and not in $\Pi$). We define a two-stage translation from
$\cGM$ into $\cC^1$ as follows.  Notice that the definition of $f_1$
makes reference to $f_2$, but not \emph{vice versa}.
\begin{align*}
f_1(p)  &= p(x) && \hspace{-1ex}\text{(for $p \in \Pi$)}\\
f_1(\phi \wedge \psi)  &= f_1(\phi) \wedge f_1(\psi) 
&&\hspace{-1ex}\text{(sim.~for $\neg$, $\vee$, etc.)}\\
f_1(\Diamond_{\geq C} \phi) &= \exists_{\geq C} .x (f_2(\phi) \wedge q_1(x))\\
f_1(\Diamond_{\leq C} \phi) &= \exists_{\leq C} x. (f_2(\phi) \wedge q_1(x))\\[2ex]
f_2(p)  &= p(x) &&\hspace{-1ex}\text{(for $p \in \Pi$)}\\
f_2(\phi \wedge \psi)  &= f_2(\phi) \wedge f_2(\psi) 
&&\hspace{-1ex}\text{(sim.~for $\neg$, $\vee$, etc.)}\\
f_2(\Diamond_{\geq C} \phi)  &= \exists_{\geq C} x. (f_2(\phi) \wedge q_2(x))\\
f_2(\Diamond_{\leq C} \phi)  &= \exists_{\leq C} x. (f_2(\phi) \wedge q_2(x)).
\end{align*}

Next, we define first-order formulas (in fact, $\cC^1$-formulas),
which, for Euclidean frames, act as substitutes for the conditions of
reflexivity, seriality, symmetry and transitivity:
\begin{align*}
\varepsilon_{\mbox{\rm \tiny Rfl}} &= \forall x. (q_0(x) \rightarrow q_1(x)) \\
\varepsilon_{\mbox{\rm \tiny Ser}} &= \exists x. q_1(x) \\
\varepsilon_{\mbox{\rm \tiny Sym}} &= \forall x. (q_0(x) \rightarrow q_1(x)) \vee 
                                \neg \exists x. q_1(x)\\
\varepsilon_{\mbox{\rm \tiny Tr}} &= \forall x. (q_2(x) \rightarrow q_1(x)).
\end{align*}
Let us define the required $\cC^1$ formula $\alpha$ as follows:
\begin{equation*}
\alpha = \exists x. (f_1(\phi) \wedge q_0(x)) \wedge
            \forall x. (q_1(x) \rightarrow q_2(x)) \wedge
              \bigwedge_{\cK \in \cF} \varepsilon_\cK.
\end{equation*}
Clearly, $\alpha$ can be constructed in polynomial time from $\phi$.  It remains
to demonstrate that $\phi$ is satisfiable over a frame in $\bigcap \cF \cap
\mbox{\rm Eucl}$ if and only if $\alpha$ is satisfiable.

Suppose $\fA \models_{w_0} \phi$, where $\fA$ is a structure over a
Euclidean frame $(W,R)$. Let $\fB$ be the substructure generated by
$\{w_0\}$---in other words, the restriction of $\fA$ to $R^*(w_0)$. By
Lemma~\ref{lma:generated}, $\fB \models_{w_0} \phi$. Expand $\fB$ to a
structure $\fB^+$ by setting
\begin{equation*}
\begin{array}{lll}
q_0^{\fB^+} = \{w_0\}, &
q_1^{\fB^+} = R(w_0), &
q_2^{\fB^+} = R(R(w_0)).
\end{array}
\end{equation*}

We shall show that $\fB^+ \models \alpha$.  By Statement~1 of
Lemma~\ref{lma:totalOrCone}, $\fB^+ \models \forall x. (q_1(x)\rightarrow
q_2(x))$.  Using Lemma~\ref{lma:totalOrCone}, a structural induction on $\psi$
easily establishes the following condition.
\begin{multline}
\text{For all $w \in q_2^{\fB^+}$, and all
$\cGM$-formulas $\psi$,}\\ \fB \models_w \psi \text{ if and only if }
\fB^+ \models f_2(\psi)[w].
\label{eq:simpleInduction1}
\end{multline}
Using~\eqref{eq:simpleInduction1}, a further structural induction establishes the 
following condition.
\begin{multline}
\text{For all $\cGM$-formulas $\psi$, }\\
   \fB \models_{w_0} \psi \text{ if and only if } \fB^+ \models f_1(\psi)[w_0].
\label{eq:simpleInduction2}
\end{multline}

From~\eqref{eq:simpleInduction2}, it follows that $\fB^+ \models
\exists x(f_1(\phi) \wedge q_0(x))$. It remains to show that, for all
$\cK \in \{\mbox{\rm Rfl}, \mbox{\rm Ser}, \mbox{\rm Sym}, \mbox{\rm
Tr} \}$, $(W,R) \in \cK$ implies $\fB^+ \models
\varepsilon_\cK$. Suppose, then $(W,R) \in \cK$; we consider the four
cases in turn.
\begin{enumerate}
\item If $\cK = \mbox{\rm Rfl}$, then $w_0 \in R(w_0)$. It follows that \newline
$\fB^+ \models \forall x. (q_0(x) \rightarrow q_1(x))$.
\item If $\cK = \mbox{\rm Ser}$, then $R(w_0) \neq \emptyset$. It follows that \newline
$\fB^+ \models \exists x. q_1(x)$.
\item If $\cK =\mbox{\rm Sym}$, then, since $(W,R)$ is both symmetric
and Euclidean, either $\langle w_0, w_0 \rangle \in R$, or \linebreak
$R(w_0) =
\emptyset$. Thus, either $\fB^+ \models \forall x. (q_0(x) \rightarrow
q_1(x))$, or $\fB^+ \models \forall x. \neg q_1(x)$.
\item If $\cK = \mbox{\rm Tr}$, then $R(R(w_0)) \subseteq R(w_0)$.
It follows that  \newline
$\fB^+ \models \forall x. (q_2(x) \rightarrow q_1(x))$.
\end{enumerate}
This establishes that $\fB^+ \models \alpha$, as required.

Conversely, suppose $\fA \models \alpha$, where $\fA$ interprets
$\Sigma$ together with the predicates $q_0$, $q_1$ and $q_2$. Let
$\fB^+$ be the substructure of $\fA$ with domain $W = q_0^\fA \cup
q_1^\fA \cup q_2^\fA$, and let $w_0 \in W$ be some element satisfying
$f_1(\phi) \wedge q_0(x)$. Since all quantification in $f_1(\phi)$ is
limited to elements satisfying $q_1$ or $q_2$, $\fB^+ \models \alpha$;
and since $\alpha$ contains no occurrences of $r$, we may without loss
of generality assume that
\begin{equation}
r^{\fB^+} = (q_0^{\fB^+} \times q_1^{\fB^+}) \cup (q_2^{\fB^+} \times q_2^{\fB^+}).
\label{eq:r}
\end{equation}
Let $\fB$ be the $\Sigma$-reduct of $\fB^+$ obtained by ignoring the predicates
$q_0$, $q_1$ and $q_2$; and let $R = r^{\fB^+}$, so that $\fB$ is a structure
over the frame $(W,R)$.  We show that $\fB \models_{w_0} \phi$, and, moreover,
$(W,R) \in \bigcap \cF \cap \mbox{Eucl}$.  Using the definition of $r^{\fB^+}$
in~\eqref{eq:r}, two simple structural inductions again
establish~\eqref{eq:simpleInduction1}, and thence~\eqref{eq:simpleInduction2}.
And from~\eqref{eq:simpleInduction2}, it follows that $\fB \models_{w_0}
\phi$. It remains to show that, for all $\cK \in \{\mbox{\rm Rfl}, \mbox{\rm
  Ser}, \mbox{\rm Sym}, \mbox{\rm Tr} \}$, $\fB^+ \models \varepsilon_\cK$
implies $(W,R) \in \cK$.  Suppose, then $\fB^+ \models \varepsilon_\cK$; we
consider the four cases in turn, making implicit use of~\eqref{eq:r} throughout.
Note also that, since $\fB^+ \models \alpha$, $q_1^{\fB^+} \subseteq
q_2^{\fB^+}$.
\begin{enumerate}
\item If $\cK = \mbox{\rm Rfl}$, $q_0^{\fB^+} \subseteq q_1^{\fB^+} \subseteq
  q_2^{\fB^+}$, whence $(W,R)$ is total, and hence certainly reflexive.
\item If $\cK = \mbox{\rm Ser}$, then $q_1^{\fB^+} \neq \emptyset$, whence
  $(W,R)$ is visibly serial.
\item If $\cK =\mbox{\rm Sym}$, either $q_0^{\fB^+} \subseteq q_1^{\fB^+}
  \subseteq q_2^{\fB^+}$ or $q_1^{\fB^+} = \emptyset$. In the former case,
  $(W,R)$ is total, and hence certainly symmetric; in the latter, $(W,R)$ is
  visibly symmetric.
\item If $\cK = \mbox{\rm Tr}$, then $q_2^{\fB^+} \subseteq q_1^{\fB^+}$,
whence $(W,R)$ is visibly transitive.
\end{enumerate}
\end{IEEEproof}

The upper bound of Theorem~\ref{theo:main1} now follows by
Theorem~\ref{theo:C1} and Lemma~\ref{lma:c1:translation}, since
$\mbox{\rm Sym} \cap \mbox{\rm Tr} \subseteq \mbox{\rm Eucl}$.  The
lower bound is trivial, since $\cGM$ includes propositional logic.

\section{Transitive frames}
\label{sec:transitive}

The purpose of this section is to establish Theorem~\ref{theo:main2}.  The upper
bound (Section~\ref{sec:upper}) is obtained by proving that every $\cGM$-formula
$\phi$ that is satisfiable over a transitive (transitive and reflexive) frame is
also satisfiable over a transitive (transitive and reflexive) frame whose size
is bounded by an exponential function of $\sizeof{\phi}$. It is shown
in~\cite{logic:ksz07} that every $\cGM$-formula satisfiable over a transitive
frame is also satisfiable over a \emph{finite} transitive frame.  However, this
paper gives no bound on the size of the satisfying
structure.  The matching
lower bound (Section~\ref{sec:lower}) is obtained by a reduction from
exponential tiling problems. Interestingly, this reduction features only
formulas in which all numerical subscripts are bounded by 1. Thus, the lower
complexity-bound of Theorem~\ref{theo:main2} continues to hold even under unary
coding of numerical subscripts.

One note on terminology before we proceed. In the context of (graded)
modal logic, it is customary to think of the
unary predicates in $\Pi$ as \emph{proposition letters}, because they
receive truth-values relative to worlds.  Since we shall not be
concerned with $\cC^1$ or other first-order fragments in the sequel,
we adopt this practice from now on. Accordingly, a \emph{propositional}
formula is one containing no modal operators.  Finally, we shall relax
our stance on valuations, allowing structures to interpret only those
proposition letters involved in some collection of formulas of
interest, rather than every proposition letter in $\Pi$.

\subsection{Membership in $\NEXPTIME$}
\label{sec:upper}
First we demonstrate that every $\cGM$-formula can be transformed into a normal
form preserving satisfiability over transitive frames. This normal form is
broadly similar to the so-called Scott normal form for the two-variable fragment
of first-order logic, and is likewise obtained by a straightforward renaming
procedure.  For the next lemma, recall that $\boxdot \phi$ abbreviates $\phi
\wedge \Box \phi$.

\begin{lemma}\label{lma:normalformGM-Tr} 
  Let $\phi$ be a $\cGM$-formula.  We can compute, in time bounded by a
  polynomial function of $\sizeof{\phi}$, a $\cGM$-formula $\psi$ of the form
\begin{equation}\label{eq:normalformGM-Tr}
\eta \wedge \boxdot \big( \theta \wedge \!\!
  \bigwedge_{1 \leq i \leq\ell}\!\! (p_i \rightarrow \Diamond_{\geq C_i} \pi_i) \wedge\!
  \! \bigwedge_{1 \leq j \leq m}\!\! (q_j \rightarrow \Diamond_{\leq D_j} \chi_j) \big),
\end{equation}
\ignore{%
\begin{equation}\label{eq:normalformGM-Tr}
\begin{split}
\eta \wedge \boxdot \big( \theta \wedge 
  \bigwedge_{1 \leq i \leq\ell} (p_i & \rightarrow \Diamond_{\geq C_i} \pi_i) \wedge\\[-2ex]
&   \bigwedge_{1 \leq j \leq m} (q_j \rightarrow \Diamond_{\leq D_j} \chi_j) \big),
\end{split}
\end{equation}
}%
where the $p_i$ and the $q_j$ are proposition letters, the $C_i$ and
$D_j$ are natural numbers, and $\eta$, $\theta$, the $\pi_i$ and the
$\chi_j$ are propositional formulas, such that $\phi$ and $\psi$ are
satisfiable over exactly the same transitive frames.
\end{lemma}

\begin{IEEEproof}
As usual, if $\rho$ is a subformula of $\phi$ and $\sigma$ a formula,
we denote by $\phi[\sigma/\rho]$ the result of substituting $\sigma$
for every occurrence of $\rho$ in $\phi$. If $\rho$ is a formula of
the form $\Diamond_{\leq C} \pi$, denote by $\bar{\rho}$ the
corresponding formula $\Diamond_{\geq (C+1)} \pi$; similarly, if
$\rho$ is a formula of the form $\Diamond_{\geq C} \pi$, with $C>0$,
denote by $\bar{\rho}$ the corresponding formula $\Diamond_{\leq
(C-1)} \pi$.

We may assume that $\phi$ contains no subformulas of the form $\Diamond_{\geq 0}
\pi$, since these may be replaced with any tautology. Suppose $\phi$ is not
propositional, and let $\rho$ be any subformula of $\phi$ having either of the
forms $\Diamond_{\leq C}\pi$ or $\Diamond_{\geq C}\pi$, with $\pi$
propositional.  (In the latter case, $C >0$.) Let $p$ and $q$ be fresh
proposition letters, and let $\phi'$ be the formula
\begin{equation*}
\phi[p/\rho] \wedge \boxdot(p \vee q) \wedge \boxdot(p \rightarrow \rho)
\wedge \boxdot(q \rightarrow \bar{\rho}).
\end{equation*}
It is easy to verify that, if $\fA \models_w \phi'$ with $\fA$
transitive, then $\fA \models_w \phi$ .  Conversely, if $\fA
\models_{w_0} \phi$, we may expand $\fA$ to a structure $\fA'$ by
setting $\fA' \models_w p$ if and only if $\fA' \models_w \rho$ and
$\fA' \models_w q$ if and only if $\fA' \not \models_w \rho$, for all
worlds $w$: evidently, $\fA' \models_{w_0} \phi'$. Thus, $\phi$ and
$\phi'$ are satisfiable over the same transitive frames. Repeating
this process and re-grouping conjuncts eventually leads to a formula
of the form~\eqref{eq:normalformGM-Tr} as required.
\end{IEEEproof}

We next present lemmas describing transformations of transitive structures, in
which we use the following terminology.  Let $\fA = \tuple{W,R,V}$ be a
transitive structure, and $w_1,w_2$ be worlds of $W$. We say: $w_2$ is an
\emph{$R$-successor of $w_1$} if $\tuple{w_1,w_2}\in R$; $w_2$ is a \emph{strict
  $R$-successor of} $w_1$ if $\tuple{w_1,w_2}\in R$, but $\tuple{w_2,w_1} \not
\in R$; $w_1$ and $w_2$ are $R$-\emph{equivalent} if $\tuple{w_1,w_2}\in R$ and
$\tuple{w_2,w_1}\in R$.  The \emph{$R$-clique for} $w_1$ in $\fA$ is the set
$Q_\fA(w_1)\subseteq W$ consisting of $w_1$ and all worlds $R$-equivalent to
$w_1$. We say that $w_2$ is a \emph{direct $R$-successor of} $w_1$ if $w_2$ is a
strict $R$-successor of $w_1$ and, for every $w\in W$ such that $\tuple{w_1,w}
\in R$ and $\tuple{w,w_2} \in R$, we have either $w \in Q_\fA(w_1)$
or $w \in Q_\fA(w_2)$.

The \emph{depth of a structure $\fA$} is the maximum over all $k \geq 0$ for
which there exist worlds $w_0,\dots,w_k\in W$ such that $w_i$ is a
strict $R$-successor of $w_{i-1}$ for every $i$ with $1\le i\le k$, or
$\infty$ if no such maximum exists.  The \emph{breadth of $\fA$} is
the maximum over all $k \geq 0$ for which there exist worlds
$w,w_1,\dots,w_k$ such that $w_i$ is a direct $R$-successor of $w$ for
every $i$ with $1\le i\le k$, and the sets
$Q_\fA(w_1),\dots,Q_\fA(w_k)$ are disjoint, or $\infty$ if no such
maximum exists.  The \emph{width of $\fA$} is the smallest $k$ such
that $k \geq\sizeof{Q_\fA(w)}$ for all $w\in W$, or $\infty$ if no such $k$
exists. 
\begin{lemma}\label{lma:newLemma}
  Let $\fA$ be a structure of depth $d$, breadth $b$ and width $c$
  \textup{(}all finite\textup{)}, and let $w$ be a world of
  $\fA$. Then the substructure of $\fA$ generated by $\{w\}$ contains
  no more than $n$ worlds, where 
$n = c$ if $b = 0$, $n = c\cdot(d+1)$ if $b = 1$, and
$n = c\cdot(b^{d+1}-1)/(b-1)$ otherwise.
%
\end{lemma}

\begin{IEEEproof}
Elementary.
\end{IEEEproof}

We employ the following notation. For a structure $\fA=(W,R,V)$ and a
binary relation $R'$ on $W$ (possibly different from $R$), we denote
by $R'_\fA(w,\phi)$ the set $\set{v\mid {\tuple{w,v}\in R'},
  {\fA\models_v\varphi}}$. Thus, $\fA \models_w \Diamond_{\geq C}
\phi$ if and only if $\sizeof{R_\fA(w,\phi)}\ge C$, where $\sizeof{S}$
denotes the cardinality of the set $S$.  Similarly, $\fA \models_w
\Diamond_{\leq C} \phi$ if and only if $\sizeof{R_\fA(w,\phi)}\leq C$.
\begin{lemma}\label{lemma:model:transformation}
  Let $\phi$ be a formula of the form~\eqref{eq:normalformGM-Tr}.  If
  $\phi$ has a transitive model $\fA$, then it has a transitive model
  $\fA'$ with depth $d' \le 2\ell$, breadth $b' \le \sum_{i=1}^\ell
  C_i$ and width $c' \le \sum_{i=1}^\ell C_i+1$. If\/~$\fA$ is
  reflexive, then we can additionally ensure that $\fA'$ is also
  reflexive.
\end{lemma}

\begin{IEEEproof}
  Let $\fA = (W,R,V)$. We construct $\fA'=(W',R',V')$ from $\fA$ in four stages.

  \noindent{\bf Stage 1: } Adapting a technique employed in~\cite{logic:ksz07}
  to establish the finite model property for $\cGM$-formulas, we first define a
  transitive model $\fA'$ of $\phi$, reflexive if $\fA$ is, such that $\fA'$ has
  finite depth.  The strategy is to {\em enlarge} the relation $R$ (thus
  reducing the number of \emph{strict} successors of worlds in $W$), preserving
  satisfaction for subformulas of the form $\Diamond_{\le D_j}\chi_j$.
  
  For $w\in W$ define $d_\fA^j(w):=\min
  (D_j+1,\sizeof{R^*(w,\chi_j)})$ where $D_j$ and $\chi_j$ $(1\le j\le
  m)$ are as in \eqref{eq:normalformGM-Tr}, and $R^*$ is the reflexive
  closure of $R$.  Let $R_d:=\set{\tuple{w_1,w_2}\in R\mid
    d_\fA^j(w_1)= d_\fA^j(w_2),\,1\le j\le m}$ be the restriction of
  $R$ to pairs of elements that have the same values of $d_\fA^j(w)$,
  and let $R_d^-:=\set{\tuple{w_1,w_2}\mid\tuple{w_2,w_1}\in R_d}$ be
  the inverse of $R_d$. Let $\fA'=(W,R',V)$ be obtained from
  $\fA=(W,R,V)$ by setting $R':=(R \cup R^-_d)^+$.  Intuitively, if
  $w_1$ is $R$-reachable from $w_2$, and, for all $j$ ($1 \leq j \leq
  m$), $w_1$ and $w_2$ agree on the number (up to the limit of $D_j$)
  of $\chi_j$-worlds that are $R$-reachable from them, then we make
  $w_1$ and $w_2$ $R'$-equivalent.  We show that $\fA'$ satisfies
  $\phi$, is reflexive if $\fA$ is, and has finite depth.

  Since $R\subseteq R'$, $\fA'$ is reflexive if $\fA$ is.  We
  claim that $\fA'$ has finite depth. Indeed, for every $w_1,w_2\in W$ such that
  $w_2$ is a strict $R'$-successor of $w_1$, we have $d_\fA^j(w_1) \ge
  d_{\fA}^j(w_2)$ for all $j$, and $d_\fA^j(w_1) > d_{\fA}^j(w_2)$ for some $j$
  $(1\le j\le m)$. Hence $\sum_{j=1}^m d_{\fA}^j(w_1)>\sum_{j=1}^m
  d_{\fA}^j(w_2)$. Since $d_{\fA}^j(w)\le D_j+1$ for every $w\in W$ and every
  $j$ $(1\le j\le m)$, the length of every chain $w_0,\dots,w_k$ such that $w_i$
  is a strict $R'$-successor of $w_{i-1}$ \textup{(}$1\le i\le k$\textup{)}, is
  bounded by $\sum_{j=1}^mD_j+m$.

  In order to prove that $\fA'$ satisfies $\phi$, we first prove that
  $d_\fA^j(w) = d_{\fA'}^j(w)$ for every $w\in W$ and every $j$ ${(1\le j\le m)}$.
  Assume to the contrary that $d_\fA^j(w) \neq d_{\fA'}^j(w)$ for some $w\in W$
  and some $j$ $(1\le j\le m)$.  Since $R\subseteq R'$, we have $d_\fA^j(w) <
  d_{\fA'}^j(w) \le D_j + 1$, which means, in particular, that there exists an
  element $w'\in W$ with $\fA\models_{w'}\chi_j$ such that $\tuple{w,w'}\in R'$
  but $\tuple{w,w'}\not\in R$.

  Since $\tuple{w,w'}\in R'$, by definition of $R'$, there exists a sequence
  $w_0,\dots,w_k$ of different worlds in $W$ such that $w_0=w$, $w_k=w'$, and
  $\tuple{w_{i-1},w_i}\in R\cup R^-_d$ for every $i$ $(1\le i\le k)$.  Note that
  $d_\fA^j(w_{i-1})\ge d_\fA^j(w_i)$ for every $i$ $(1\le i\le k)$ and every $j$
  $(1\le j\le m)$. Take the maximal $i$ $(1\le i \le k)$ such that
  $\tuple{w_{i-1},w'}\notin R$. Since $\tuple{w_0,w'}=\tuple{w,w'}\notin R$,
  such a maximal $i$ always exists. Then $\tuple{w_i,w'}\in R^*$, and
  $\tuple{w_{i-1},w_i}\notin R$.  Since $\tuple{w_{i-1},w_i}\in R\cup R^-_d$, we
  have $\tuple{w_{i-1},w_{i}}\in R^-_d$, and so $d_\fA^j(w_{i-1})= d_\fA^j(w_i)$
  by definition of $R_d$. Since $d_\fA^j(w_i)\le d_\fA^j(w_0)=d_\fA^j(w)<
  D_j+1$, we obtain a contradiction, due to the fact that
  $d_\fA^j(w_{i-1})=d_\fA^j(w_{i})\le D_j$, $\tuple{w_{i-1},w'}\notin R^*$,
  $\tuple{w_i,w'}\in R^*$, and $\fA\models_{w'}\chi_j$.
 
  Now to complete the proof that $\fA'$ satisfies $\phi$, we
  demonstrate that, if $\psi$ is any of the formulas $\eta$, $\theta$,
  $(p_i\rightarrow\Diamond_{C_i}\pi)$ or
  $(q_j\rightarrow\Diamond_{\leq D_j}\chi_j)$ occurring
  in~\eqref{eq:normalformGM-Tr}, and $w\in W$, then $\fA\models_w\psi$
  implies $\fA'\models_w\psi$. Indeed, for the propositional subformulas
  $\eta$ and $\theta$, this is immediate. For
  subformulas $p_i\rightarrow\Diamond_{\ge C_i}\pi_i$, this holds
  since $R\subseteq R'$. Finally, for subformulas
  $q_j\rightarrow\Diamond_{\le D_j}\chi_j$ this follows from the
  property $d_\fA^j(w) = d_{\fA'}^j(w)$.

  \noindent{\bf Stage 2: } By Stage~1, we may assume that $\fA$ has finite depth
  $d$.  We define a transitive model $\fA'$ of $\phi$, reflexive if $\fA$ is,
  such that $\fA'$ has depth $d' \leq 2\ell$. If $d\leq 2\ell$ then we take
  $\fA'=\fA$. Otherwise, we obtain $\fA'$ from $\fA$ by \emph{contracting} the
  relation $R$ (removing unnecessary \emph{direct} successors of worlds in $W$),
  preserving satisfaction for subformulas of the form $\Diamond_{\ge
    C_i}\pi_i$. Define, for every $w \in W$, two sets of indices:
  \begin{align*}
    I_\fA(w)&=\set{i \mid {1\le i\le
        \ell},  \sizeof{R(w,\pi_i)}\ge C_i},\text{ and}\\
    I_\fA^s(w)&=\set{i \mid {1\le
        i\le\ell},  \sizeof{R(w,\pi_i)\setminus Q_{\fA}(w)}\ge C_i},
  \end{align*}
where $\pi_i$ and $C_i$ are as in \eqref{eq:normalformGM-Tr}, $1\le
i\le \ell$.  Note that: 
 \begin{itemize}
 \item[({\bf P1})] $I_\fA^s(w)\subseteq I_\fA(w)$ for every $w\in W$, and
 \item[({\bf P2})] $I_\fA(w_2)\subseteq I_\fA^s(w_1)$ if $w_2$ is a strict
   $R$-successor of $w_1$.
 \end{itemize}
Define the structure $\fA' = \tuple{W,R',V}$ by setting
  \begin{equation*}
\begin{split}
R':=R\setminus \set{\tuple{w_1,w_2}\mid {}
          & w_2\text{ is a direct $R'$-successor of }w_1\\
          & \text{ and }I_{\fA}^s(w_2)=I_{\fA}(w_1)}.
\end{split}
 \end{equation*}
 We claim that $\fA'$ is a transitive structure which satisfies $\varphi$,
 is reflexive if $\fA$ is, and has depth $d' < d$.  Repeating this step
 sufficiently often, we eventually ensure that $d' \le 2\ell$.

 It is easy to see that $R'$ is transitive if $R$ is transitive. Indeed, if
 $\tuple{w_1,w_2}\in R'$ and $\tuple{w_2,w_3}\in R'$, we have
 $\tuple{w_1,w_3}\in R$, and either $(i)$~$w_3$ is not a direct $R$-successor of
 $w_1$, or $(ii)$~$w_2\in Q_\fA(w_1)$ and $I_\fA^s(w_3)\neq
 I_\fA(w_2)=I_\fA(w_1)$, or $(iii)$~$w_2\in Q_\fA(w_3)$ and $I_\fA^s(w_3)=
 I_\fA^s(w_2)\neq I_\fA(w_1)$. In all of these three cases, we have
 $\tuple{w_1,w_3}\in R'$ by the definition of $R'$.  Trivially, $R'$ is
 reflexive if $R$ is.

 In order to prove that $\fA'$ satisfies $\phi$, we first point out some other
 properties of $I_\fA(w)$, $I_\fA^s(w)$, $I_{\fA'}(w)$, and $I_{\fA'}^s(w)$:

 \begin{itemize}
 \item[({\bf P3})]~$I_{\fA'}(w)\subseteq I_\fA(w)$ and
   $I_{\fA'}^s(w)\subseteq I_\fA^s(w)$ for $w\in W$;
 \item[({\bf P4})]~$I_\fA^s(w_2)\subseteq I_{\fA'}(w_1)$ if $w_2$ is a strict $R$-successor of $w_1$;
 \item[({\bf P5})]~$I_{\fA'}(w)= I_\fA(w)$ for $w\in W$.
 \end{itemize}
 
 Property ({\bf P3}) holds since $R'\subseteq R$.  Property ({\bf P4}) holds since, for every
 $i$ $(1\le i\le\ell)$, every $w_3\in R_\fA(w_2,\pi_i)\setminus Q_\fA(w_2)$ is a
 strict non-direct $R$-successor of $w_1$.  Hence $\tuple{w_1,w_3}\in R'$ by the
 definition of $R'$, and so, $w_3\in R_{\fA'}(w_1,\pi_i)$. In order to prove
 ({\bf P5}), by ({\bf P3}), it suffices to prove $I_{\fA'}(w) \supseteq
 I_\fA(w)$. Assume to the contrary that there exists $w\in W$ and $i$ $(1\le
 i\le\ell)$ such that $\fA \models_{w'} \pi_i$ (equivalently, $\fA' \models_{w'}
 \pi_i$), $\tuple{w,w'}\in R$, and $\tuple{w,w'} \notin R'$. By the definition
 of $R'$, this is only possible if $w'$ is a direct $R$-successor of $w$ and
 $I_\fA^s(w')=I_\fA(w)$. But then, by ({\bf P4}), we have $I_\fA^s(w')\subseteq
 I_{\fA'}(w)$. Hence $I_\fA(w)=I_\fA^s(w')\subseteq I_{\fA'}(w)$, which
 contradicts the assumption that $I_\fA(w)\setminus
 I_{\fA'}(w)\neq\emptyset$.

 In order to prove that $\fA'$ satisfies $\phi$, it is sufficient, as in
 Stage~1, to demonstrate that, if $\psi$ is any of the formulas $\eta$,
 $\theta$, $(p_i\rightarrow\Diamond_{C_i}\pi_i)$ or
 $(q_j\rightarrow\Diamond_{\leq D_j}\chi_j)$ occurring
 in~\eqref{eq:normalformGM-Tr}, and $w\in W$, then $\fA\models_w\psi$ implies
 $\fA'\models_w\psi$. This property holds for $\psi=\eta$, $\psi=\theta$, and
 $\psi={(q_j\rightarrow\Diamond_{\leq D_j}\chi_j)}$, $1\le j\le m$, since 
 $R'\subseteq R$. For $\psi={(p_i\rightarrow\Diamond_{C_i}\pi_i)}$, $1\le
 i\le m$, this property holds by ({\bf P5}).
 
 Finally, it remains to demonstrate that the depth of $\fA'$ is smaller than the
 depth $d$ of $\fA$.  Suppose, to the contrary, that there exists a sequence of
 worlds $w_0,\dots,w_d$ in $W$ such that $w_i$ is a strict $R'$-successor of
 $w_{i-1}$, $1\le i\le d$. By definition of $R'$, every $w_i$ is a strict
 $R$-successor of $w_{i-1}$, and, since $d$ is the depth of $\fA$, $w_i$ is in
 fact a direct $R$-successor of $w_{i-1}$, $1\le i\le d$. Again, by definition
 of $R'$, we have $I_\fA^s(w_i)\neq I_\fA(w_{i-1})$, $1\le i\le d$. By ({\bf P1}) 
 and ({\bf P2}) 
 we have $I_\fA^s(w_i)\subsetneq I_\fA(w_{i-1})$ and $I_\fA(w_i)\subseteq
 I_\fA^s(w_{i-1})$, so
 $\sizeof{I_\fA^s(w_i)}+\sizeof{I_\fA(w_i)}<\sizeof{I_\fA^s(w_{i-1})}+\sizeof{I_\fA(w_{i-1})}$,
 $1\le i\le d$.  Since $\sizeof{I_\fA^s(w)}\le \sizeof{I_\fA(w)}\le \ell$ for
 every $w$ in $W$, this is possible only if $d\le 2\ell$.

 \noindent{\bf Stage 3: } By Stage~2, we may assume that $\fA$ has depth $d \leq
 2\ell$.  We define a transitive model $\fA'$ of $\phi$, reflexive if $\fA$ is,
 such that $\fA'$ has depth $d' \leq 2\ell$ and breadth $b' \le \sum_{i=1}^\ell
 C_i$.  For every element $w\in W$ and every $i$ with $1\le i\le\ell$, let
 $W_i(w)$ be the set of strict $R$-successors of $w$ for which $\pi_i$ holds. We
 call the elements of $W_i(w)$ the \emph{strict $\pi_i$-witnesses for $w$}. Note
 that $W_i(w_1)=W_i(w_2)$ when $w_1$ and $w_2$ are $R$-equivalent.  Let
 $W_i'(w)$ be $W_i(w)$ if $\sizeof{W_i(w)}\le C_i$ or, otherwise, a subset of
 $W_i(w)$ which contains exactly $C_i$ elements. We call $W_i'(w)$ the
 \emph{selected strict $\pi_i$-witnesses for $w$}.  We assume that
 $W_i'(w_1)=W_i'(w_2)$ when $w_1$ and $w_2$ are $R$-equivalent.  Let
 $R_q:=\set{\tuple{w,w'}\in R\mid w'\in Q_\fA(w)}$ be the restriction of $R$ to
 elements of the same clique, and $R'_i=\set{\tuple{w,w'}\in R\mid w'\in
   W_i'(w)}$ be the relation between an element $w\in W$ and the selected strict
 $\pi_i$-witnesses for $w$.  Define the structure $\fA' = (W,R',V)$ by setting
 $R':=(R_q\cup\bigcup_{1\le i\le\ell} R_i')^+$. Intuitively, $\fA'$ is obtained
 from $\fA$ by removing all strict successor relations except those that are
 induced by selected strict witnesses. We show that $\fA'$ has all required
 properties.

 Note that $R'$ is transitive, and reflexive if $R$ is reflexive.  Clearly, the
 depth of $\fA'$ is bounded by $d$, since only strict successor relations are
 removed. It is also clear that the breadth of $\fA'$ is bounded by
 $b=\sum_{i=1}^\ell C_i$, since for every $w\in W$ and every direct
 $R'$-successor $w'$ of $w$ there exists $i$ with $1\le i\le\ell$ such that
 $Q_\fA(w')\cap W_i'(w)\neq\emptyset$, and so the maximal number of such
 successors $w'$ for which $Q_\fA(w')$ are disjoint is bounded by
 $\sum_{i=1}^\ell\sizeof{W_i'(w)}\le\sum_{i=1}^\ell C_i=b$.

 It remains to demonstrate that $\fA'$ satisfies $\phi$. Clearly, the set of
 worlds $w\in W$ that satisfy subformulas $\eta$ and $\theta$ has not changed. 
 The set of worlds that satisfy subformulas
 $(q_j\rightarrow\Diamond_{\le D_j}\chi_j)$ can only have increased, since
 $R'\subseteq R$. Finally, the set of worlds that satisfy subformulas
 $(p_i\rightarrow\Diamond_{\ge C_i}\pi_i)$ has not changed, since, for every
 $w\in W$, the number of direct $\pi_i$-witnesses has either not changed, or is
 at least $C_i$.

 \noindent{\bf Stage 4: } By Stage~3, we may assume that $\fA$ has depth $d \leq
 2\ell$ and breadth $b \le \sum_{i=1}^\ell C_i$. We define a structure $\fA'$
 with all the properties required by the lemma. For every element $w\in W$, and
 every $i$ with $1\le i\le\ell$, let $Q_i(w)$ be the set of elements in
 $Q_\fA(w)$ for which $\pi_i$ holds. We call the elements of $Q_i(w)$ the
 \emph{equivalent $\pi_i$-witnesses for $w$}. Note that $Q_i(w_1)=Q_i(w_2)$ when
 $w_1$ and $w_2$ are $R$-equivalent. Let $Q_i'(w)$ be $Q_i(w)$ if
 $\sizeof{Q_i(w)}\le C_i$ or, otherwise, a subset of $Q_i(w)$ which contains
 exactly $C_i$ elements.  We call $Q_i'(w)$ the \emph{selected equivalent
   $\pi_i$-witnesses for $w$}.  Also let $Q_0'(w)$ be a singleton set containing
 an element of $Q_\fA(w)$ that satisfies $\phi$ if there is one, and any element
 of $Q_\fA(w)$ otherwise.  We assume that $Q_i'(w_1)=Q_i'(w_2)$ when $w_1$ and
 $w_2$ are $R$-equivalent.  Define the structure $\fA' = \tuple{W',R',V'}$ by
 setting $W':=\bigcup_{w\in W,\,0\le i\le\ell}Q_i'(w)$, $R':=R|_{W'}$, and
 $V':=V|_{W'}$.  Intuitively $\fA'$ is obtained from $\fA$ by removing elements
 in every $R$-clique, except for those that are selected witnesses of other
 elements, and in such a way that the clique remains non-empty and contains at
 least one element satisfying $\phi$ if there was one.  (Note that, since no
 $R$-clique is completely obliterated by this process, $W'$ is non-empty.)  We
 show that $\fA'$ has all required properties.

 Clearly, $\fA'$ is a transitive structure, and indeed is
 reflexive if $\fA$ is reflexive.  Further, the depth and breadth of $\fA'$ is
 bounded by the depth and breadth of $\fA$ since $\fA'$ is a restriction of
 $\fA$ to a subset of $W$.  It is easy to see that for every $w\in W'$,
 $Q_{\fA'}(w)=\bigcup_{0\le i\le\ell}Q_i'(w)$. Hence $\sizeof{Q_{\fA'}(w)}\le
 \sum_{i=0}^{\ell}\sizeof{Q_i'(w)}\le \sum_{i=1}^{\ell}C_i+1=c$. Therefore the
 width of $\fA'$ is bounded by $c$.

  It remains to demonstrate that $\fA'$ satisfies $\phi$. By the definition of
  $W'$ there is a world $w_0\in W'$ such that $\fA\models_{w_0}\phi$. Clearly
  $\fA'\models_{w_0}\eta$ since $\fA\models_{w_0}\eta$ and $V'=V|_{W'}$. Let
  $w\in W$ be any world such that $\tuple{w_0,w}\in R'$. We need to demonstrate
  that $(i)$~$\fA'\models_w\theta$,
  $(ii)$~$\fA'\models_w(p_i\rightarrow\Diamond_{\ge C_i}\pi_i)$, $1\le
  i\le\ell$, and $(iii)$~$\fA'\models_w(q_j\rightarrow\Diamond_{\le D_j}\chi_j)$,
  $1\le j\le m$. Cases $(i)$ and $(iii)$ are trivially satisfied since
  $V'=V|_{W'}$ and $R'\subseteq R$. Case $(ii)$ is satisfied since, for every
  $i$ with $1\le i\le \ell$, $\sizeof{R_\fA(w,\pi_i)}\ge C_i$ implies
  $\sizeof{R'_{\fA'}(w,\pi_i)}\ge C_i$.

\end{IEEEproof}

\begin{lemma}\label{lma:exp:model}
  Let $\fA = \tuple{W,R,V}$ be a transitive structure that satisfies a
  formula $\phi$ of the form~\eqref{eq:normalformGM-Tr}.  Then there
  exists a transitive structure $\fA'=\tuple{W',R',V'}$ that satisfies
  $\phi$ such that $\sizeof{W'}\le (b+1)\cdot (b^{2\ell+1}-1)/(b-1)$,
  where $b=\max(2,\sum_{i=1}^\ell C_i)$.  Moreover, if $\fA$ is reflexive,
  then we can ensure that $\fA'$ is also reflexive.
\end{lemma}

\begin{IEEEproof}
  By Lemma~\ref{lemma:model:transformation}, there is a transitive structure
  $\fA'$ satisfying $\phi$, reflexive if $\fA$ is, with depth, breadth, and
  width bounded respectively by $2\ell$, $b$, and $b+1$. Let $w_0$ be such that
  $\fA'\models_{w_0}\phi$, and consider the substructure of $\fA'$ generated
  by~$\{w_0\}$.  The result now follows by Lemmas~\ref{lma:generated}
  and~\ref{lma:newLemma}.
\end{IEEEproof}

We remark that the bound $(b+1)\cdot (b^{2\ell+1}-1)/(b-1)$ obtained in
Lemma~\ref{lma:exp:model} is at most exponential in the size of the input
formula, even under binary coding of the numerical subscripts $C_1, \ldots,
C_{\ell}$.  Notice, incidentally, that this bound does not mention the
subscripts $D_1, \ldots, D_m$ at all.

\begin{corollary}\label{corollary:membership}
If $\cF$ is any of $\{\mbox{\rm Tr}\}$, $\{\mbox{\rm
  Rfl}, \mbox{\rm \small Tr}\}$ or $\{\mbox{\rm Ser}, \mbox{\rm
  Tr}\}$, then the problem $\cGM_{\cap \cF}$-Sat is in $\NEXPTIME$.
\end{corollary}

\begin{IEEEproof}
  Consider first the cases $\cF = \{\mbox{\rm Tr}\}$ and $\cF = \{\mbox{\rm Tr},
  \mbox{\rm Rfl}\}$.  By Lemma~\ref{lma:normalformGM-Tr}, any $\cGM$ formula
  $\phi$ can be transformed in polynomial time into a formula $\psi$ of the form
  \eqref{eq:normalformGM-Tr} preserving satisfiability over $\bigcap \cF$. By
  Lemma~\ref{lma:exp:model}, $\psi$ is satisfiable over $\bigcap \cF$ if and
  only if it is satisfiable over a frame in $\bigcap \cF$ of size at most
  exponential in $\sizeof{\psi}$. This last condition can be checked in
  non-deterministic exponential time.  Finally, using Lemma~\ref{lma:generated},
  a formula $\phi$ is satisfiable over $\mbox{Ser} \cap \mbox{Tr}$ if and only
  if $\phi \wedge \boxdot \Diamond \top $ is satisfiable over $\mbox{Tr}$, where
  $\top$ is any tautology.
\end{IEEEproof}

\subsection{$\NEXPTIME$-hardness}
\label{sec:lower}
\begin{figure*}
  \subfloat[The set of all z-worlds forming a (rather jumbled) `ziggurat' under
  the direct successor relation. The world $w_0$, with character $(0,0)$,
  lies at the apex of the ziggurat, and the worlds
  with character $(n,n)$ form its base.]{%
\label{fig:z:worlds:ziggurat}%
    \raisebox{.5cm}{%
\begin{tabular}[c]{@{}c@{}}
\includegraphics{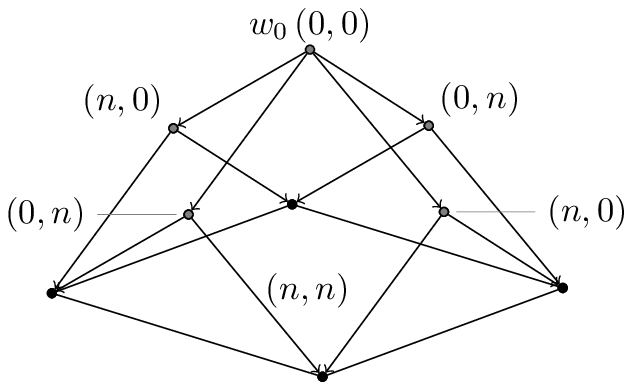}
\end{tabular}}
}\hfill%
\subfloat[The direct successors of a z-world with character $(i,j)$,
where $0 \leq i < n$ and $0 \leq j < n$. Any such z-world
has four direct successors: two with character
$(i+1,j)$ and complementary values of $p_{i+1}$, and two with
character $(i,j+1)$ and complementary values of $q_{j+1}$.]{%
\label{fig:z:worlds:successor}%
\begin{tabular}[t]{@{}c@{}}
\includegraphics{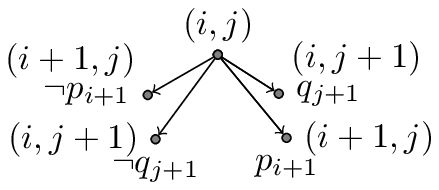}
\end{tabular}
}\hfill%
\subfloat[Identifying z-worlds with the same indices using
Formulas~\eqref{eq:character4}--\eqref{eq:character6}.  From every z-world $w$
with character $(i,j)$, we can access at most two z-worlds $a$ and $c$ with
character $(i+1,j)$, at most two z-worlds $b$ and $d$ with character $(i,j+1)$,
and at most four (not eight!) z-worlds $x$, $y$, $u$ and $v$ with character
$(i+1,j+1)$.]{%
\label{fig:z:worlds:merge}%
\begin{tabular}[t]{@{\ \quad}c@{\quad\ }}
\includegraphics{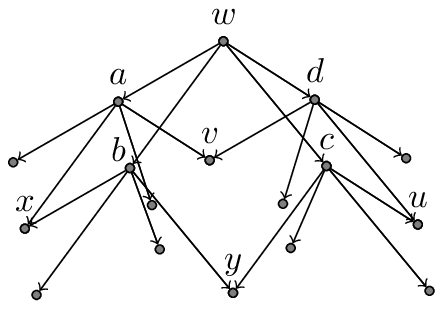}
\end{tabular}
}
\caption{The set of z-worlds generated by
  Formulas~\eqref{eq:generate0}--\eqref{eq:character6}.  
    }
\label{fig:z:worlds}
\end{figure*}

To prove a matching lower bound, we employ the apparatus of tiling systems.  A
\emph{tiling system} is a triple $\langle C, H, V \rangle$, where $C$ is a
non-empty, finite set and $H$, $V$ are binary relations on $C$. The elements of
$C$ are referred to as \emph{colours}, and the relations $H$ and $V$ as the
\emph{horizontal} and \emph{vertical} constraints, respectively. For any integer
$N$, a \emph{tiling} for $\langle C, H, V \rangle$ of size $N$ is a function $f:
\{0, \ldots, N-1\}^2 \rightarrow C$ such that, for all $i, j$ with $0\le i<N-1$,
$0\le j\le N-1$, the pair $\tuple{f(i,j), f(i+1,j)}$ is in $H$ and for all $i,
j$ with $0\le i\le N-1$, $0\le j< N-1$, the pair $\langle f(i,j), f(i,j+1)
\rangle$ is in $V$. A tiling of size $N$ is to be pictured as a colouring of an
$N \times N$ square grid by the colours in $C$; the horizontal constraints $H$
thus specify which colours may appear `to the right of' which other colours; the
vertical constraints $V$ likewise specify which colours may appear `above' which
other colours. An $n$-tuple $\bar{c}$ of elements of $C$ is an \emph{initial
  configuration} for the tiling $f$ if $\bar{c} = f(0,0), \ldots, f(n-1,0)$. An
initial configuration for $f$ is to be pictured as a row of $n$ colours
occupying the bottom left-hand corner of the grid.

Let $(C,H,V)$ be a tiling system and $p$ a polynomial. The {\em exponential
  tiling problem} $(C,H,V,p)$ is the following problem: given an $n$-tuple
$\bar{c}$ from $C$, determine whether there exists a tiling for $(C,H,V)$ of
size $2^{p(n)}$ with initial configuration $\bar{c}$.  It is well-known that
there exist exponential tiling problems which are $\NEXPTIME$-complete (see,
e.g. \cite{BoGG97a}, pp.~242, ff.). We show how, for any class of
frames $\cK$ such that $\mbox{\rm Tr} \supseteq \cK \supseteq \mbox{\rm Tr} \cap
\mbox{\rm Rfl}$, any exponential tiling problem $(C,H,V,p)$ can be reduced to
$\cGM_\cK$-Sat, in polynomial time.

In the sequel, we denote by $\{0,1\}^*$ the set of finite strings over the
alphabet $\{0,1\}$; we denote the length of any $s \in \{0,1\}^*$ by
$\sizeof{s}$; we denote the empty string by $\epsilon$; and we write $s \preceq
t$ if $s$ is a (proper or improper) prefix of $t$. If $\sizeof{s} = k$, then $s$
encodes a number in the range $[0,2^k-1]$ in the usual way; we follow standard
practice in taking the left-most digit of $s$ to be the most significant. We
equivocate freely between strings and the numbers they represent; in particular,
we write $s+1$ to denote the string representing the successor of the number
represented by $s$. Finally, if $s$ is a string and $1 \leq k \leq \sizeof{s}$,
denote the $k$th element of $s$ (counting from the left and starting with 1) by
$s[k]$. We use the notation $\pm_i\phi$ (with $i$ a numerical subscript), to
stand, ambiguously, for the formulas $\phi$ or $\neg \phi$. All occurrences of
$\pm_i\phi$ within a single formula should be expanded in all possible ways to
$\phi$ and $\neg\phi$ such that occurrences with the same index $i$ are expanded
in the same way.

We are going to write formulas that induce a structure similar to that depicted
in Fig.~\ref{fig:z:worlds:ziggurat}, the bottom of which will represent the grid
associated with (an instance of) a tiling problem.  Fix $n >0$.  We consider
structures interpreting the proposition letters $u_0,\dots,u_n$,
$v_0,\dots,v_n$, $p_1,\dots,p_n$, $q_1,\dots,q_n$, $z$, $o_h$ and $o_v$.  Let
$\Gamma_1$ be the set of all formulas:
\begin{align}
&\begin{array}{@{}l@{}}
u_0 \wedge v_0 \wedge z
\end{array}
&&\label{eq:generate0}\\
&\begin{array}{@{}l@{}}
\boxdot (\neg(u_i \wedge u_j) \wedge \neg(v_i \wedge v_j))
\end{array}
&&\begin{array}{@{}l@{}}
(0 \leq i < j \leq n)
\end{array}
\label{eq:character}\\
&\begin{array}{@{}l@{}}
\boxdot(u_i \wedge v_j \wedge z \rightarrow\\
\qquad\Diamond(u_{i+1} \wedge v_j \wedge z \wedge \pm_1 p_{i+1}))
\end{array}
&&\begin{array}{@{}l@{}}
(0 \leq i < n,\\
\phantom{(}0 \leq j \leq n)
\end{array}
\label{eq:generate1}\\   
&\begin{array}{@{}l@{}}
\boxdot(u_i \wedge v_j \wedge z \rightarrow\\
\qquad\Diamond(u_i \wedge v_{j+1} \wedge  z \wedge \pm_1 q_{j+1}))
\end{array}
&&\begin{array}{@{}l@{}}
(0 \leq i \leq n,\\
\phantom{(}0 \leq j < n)
\end{array}
\label{eq:generate2}\\
&\begin{array}{@{}l@{}}
\Box(u_i \wedge \pm_1 p_k \rightarrow \Box (z \rightarrow \pm_1 p_k))
\end{array}
&&\begin{array}{@{}l@{}}
(1 \leq k\leq i \le n)
\end{array}
\label{eq:generate3}\\
&\begin{array}{@{}l@{}}
\Box(v_j \wedge \pm_1 q_k \rightarrow\Box(z \rightarrow \pm_1 q_k))
\end{array}
&&\begin{array}{@{}l@{}}
(1 \leq k\leq j \le n)
\end{array}
\label{eq:generate4}
\end{align}

Suppose $\fA$ is a transitive structure and $w_0$ a world of
$\fA$ such that $\fA \models_{w_0} \Gamma_1$. We employ the following
terminology. A world $w$ of $\fA$ \emph{has character} $(i,j)$, for
$i, j$ in the range $[0,n]$, if $\fA \models_w u_i \wedge v_j$.  A
\emph{z-world} is a member of the smallest set $Z$ of worlds such that: $(i)$
$w_0 \in Z$; and $(ii)$ if $w \in Z$, and $w'$ is a direct successor of $w$ with
$\fA \models_{w'} z$, then $w' \in Z$. (Notice that the definition of z-world
depends on $w_0$; where $w_0$ is not clear from context, we speak of a z-world
{\em relative to} $w_0$.)  Necessarily, every z-world is either identical to, or
accessible from, $w_0$.  For any z-world $w$, with character $(i,j)$, we define
strings $s, t \in \{0,1\}^*$ of length $i$ and $j$, respectively, by setting
$s[k] = 1$ if and only if $\fA \models_w p_k$ for all $k$ ($1 \leq k \leq i$),
and $t[k] = 1$ if and only if $\fA \models_w q_k$ for all $k$ ($1 \leq k \leq
j$). The quadruple $(i,j,s,t)$ is the \emph{index} of $w$.

To see that Formulas~\eqref{eq:generate0}--\eqref{eq:generate4} generate the
structure in Fig.~\ref{fig:z:worlds:ziggurat}, note first that
Formula~\eqref{eq:generate0} implies the existence of a z-world $w_0$ with
character $(0,0)$.  Formulas~\eqref{eq:character} ensure that every z-world has
a unique character.  If $0 \leq i < n$ and $0 \leq j < n$, then
Formulas~\eqref{eq:generate1} and~\eqref{eq:generate2} imply that every z-world
with character $(i,j)$ has four direct successors: two with character $(i+1,j)$
and complementary values of $p_{i+1}$, and two with character $(i,j+1)$ and
complementary values of $q_{j+1}$ (Fig.~\ref{fig:z:worlds:successor}).
Similarly, if $0 \leq i < n$ and $j = n$, or if $0 \leq j < n$ and $i = n$,
every z-world with character $(i,j)$ has two direct successors.

\begin{lemma}\label{lma:generate1}
  Suppose $\fA \models_{w_0} \Gamma_1$.  Let $w$ be a z-world with index
  $(i,j,s,t)$, and suppose $i'$, $j'$, $s'$, $t'$ satisfy: \textup{(}i\textup{)}
  $i \leq i' \leq n$; \textup{(}ii\textup{)} $j \leq j' \leq n$;
  \textup{(}iii\textup{)} $i+j < i' + j'$; \textup{(}iv\textup{)} $s \preceq s'$
  and $\sizeof{s'} = i'$; and \textup{(}v\textup{)} $t \preceq t'$ and
  $\sizeof{t'} = j'$.  Then there exists a z-world $w'$, accessible from $w$,
  with index $(i',j',s',t')$.
\end{lemma}

\begin{IEEEproof}
  Easy induction using Formulas~\eqref{eq:generate1}--\eqref{eq:generate4}.
\end{IEEEproof}

\begin{lemma}\label{lma:generate2}
  Suppose $\fA \models_{w_0} \Gamma_1$.  For all $i$ \textup{(}$0 \leq i \leq
  n$\textup{)}, all $j$ \textup{(}$0 \leq j \leq n$\textup{)}, all $s \in
  \{0,1\}^*$ \textup{(}$\sizeof{s} = i$\textup{)} and all $t \in \{0,1\}^*$
  \textup{(}$\sizeof{t} = j$\textup{)}, there exists a z-world with index
  $(i,j,s,t)$.
\end{lemma}

\begin{IEEEproof}
  From Lemma~\ref{lma:generate1} and the fact that $w_0$ has index
  $(0,0,\epsilon,\epsilon)$.
\end{IEEEproof}

We now add formulas limiting the number of z-worlds with any given character
(see Fig.~\ref{fig:z:worlds:merge}). In particular, z-worlds will turn out to
be uniquely identified by their indices.  Let $\Gamma_2$ be the set of formulas:
\begin{align}
&\begin{array}{@{}l@{}}
\boxdot(u_i \wedge v_j \rightarrow\\
\qquad\Diamond_{\leq 1} (u_{i+1} \wedge v_j \wedge \pm_1 p_{i+1}))
\end{array}
&&\begin{array}{@{}l@{}}
(0 \leq i < n,\\
\phantom{(} 0 \leq j \leq n)
\end{array}
\label{eq:character4}\\
&\begin{array}{@{}l@{}}
\boxdot(u_i \wedge v_j \rightarrow\\
\qquad\Diamond_{\leq 1} (u_i \wedge v_{j+1} \wedge \pm_1 q_{j+1}))
\end{array}
&&\begin{array}{@{}l@{}}
(0 \leq i \leq n,\\
\phantom{(} 0 \leq j < n)
\end{array}
\label{eq:character5}\\
&\begin{array}{@{}l@{}}
\boxdot (u_i \wedge v_j \rightarrow\\
\qquad\Diamond_{\leq 1}(u_{i+1} \wedge v_{j+1} \wedge\\
\qquad\qquad\pm_1 p_{i+1} \wedge \pm_2 q_{j+1}))
\end{array}
&&\begin{array}{@{}l@{}}
(0 \leq i < n,\\
\phantom{(} 0 \leq j < n)
\end{array}
\label{eq:character6}
\end{align}

\begin{figure*}
\subfloat[][The ziggurat, together with the grid at its base.]{%
\label{fig:o:worlds:ziggurat}%
\begin{tabular}{@{}c@{}}
\includegraphics{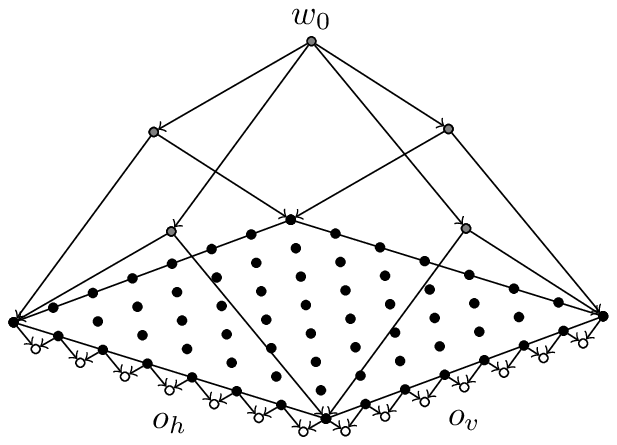}
\end{tabular}
}\hfill \subfloat[][The world arrangement for the grid.]{%
\label{fig:o:worlds:grid}%
\begin{tabular}{@{}c@{}}
\includegraphics{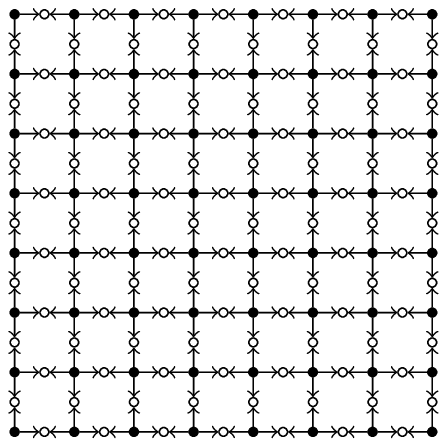}
\end{tabular}
}\hfill \subfloat[][An illustration of Formulas~\eqref{eq:standard3:h} and
Lemma~\ref{lma:standard:gamma1}.]{%
\begin{tabular}{@{}c@{}}
\label{fig:o:worlds:merge}%
\includegraphics{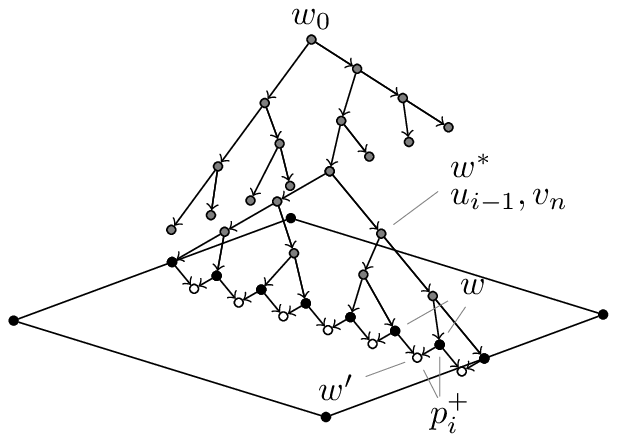}
\end{tabular}
}%
\caption{Creating o-worlds (shown as a hollow dots) and the grid using
  Formulas~\eqref{eq:standard1:h}--\eqref{eq:standard3:v} ($n=3$):
  g-worlds (shown as filled dots) are arranged according to their
  coordinates at the base; 
  g-worlds which are horizontal neighbours in this grid
  have a common horizontal o-world successor, while
  g-worlds which are vertical neighbours in this grid
  have a common vertical o-world successor.}
\label{fig:o:worlds}
\end{figure*}

\begin{lemma}\label{lma:uniqueIndex}
  Suppose $\fA \models_{w_0} \Gamma_1 \cup \Gamma_2$.  Then no two different
  z-worlds have the same index.
\end{lemma}
\begin{IEEEproof}
  Order the pairs of integers in the range $[0,n]$ in some way such that $i+j <
  i' + j'$ implies $(i,j) < (i',j')$, and proceed by induction on the character
  $(i,j)$ of z-worlds, under this ordering.

  \smallskip

  \noindent
  {\bf Case 1:} $w$ has character $(0,0)$.  By definition, $w_0$
  is the only z-world with character $(0,0)$, and hence the only
  z-world with index $(0,0, \epsilon,\epsilon)$.

  \smallskip

  \noindent
  {\bf Case 2:} $w_1$ and $w_2$ have index $(i+1,j+1,sa,tb)$ where, $0 \leq i <
  n$, $0 \leq j < n$ and $a, b \in \{0,1\}$. If $w_1$ and $w_2$ are z-worlds,
  there exist z-worlds $w'_1$ and $w'_2$ such that $w_i$ is a direct successor
  of $w'_i$ ($1 \leq i \leq 2$).  The possible characters of $w'_1$ and $w'_2$
  are $(i+1,j)$ and $(i,j+1)$. If $w'_1$ and $w'_2$ have the same character,
  then they in fact have the same index (this follows from
  Formulas~\eqref{eq:generate3} and~\eqref{eq:generate4}, and the fact that
  $w_1$ and $w_2$ have the same index).  By inductive hypothesis, then, $w'_1 =
  w'_2$. Hence, from Formulas~\eqref{eq:character4} or~\eqref{eq:character5},
  $w_1 = w_2$ as required. If $w'_1$ and $w'_2$ have different characters,
  assume without loss of generality that $w'_1$ has index $(i,j+1,s,tb)$, and
  $w'_2$ has index $(i+1,j,sa,t)$. By Lemma~\ref{lma:generate2}, let $w^*$ be
  any z-world with index $(i,j,s,t)$. By Lemma~\ref{lma:generate1}, let $w''_1$
  and $w''_2$ be z-worlds, accessible from $w^*$, with indices $(i,j+1,s,tb)$,
  and $(i+1,j,sa,t)$, respectively. By inductive hypothesis, $w'_1 = w''_1$, and
  $w'_2 = w''_2$: that is to say, $w'_1$ and $w'_2$ are accessible from $w^*$.
  Therefore, so are $w_1$ and $w_2$. Formulas~\eqref{eq:character6} then ensure
  that $w_1 = w_2$.

  \smallskip

  \noindent
  {\bf Case 3:} $w_1$ and $w_2$ have index $(i+1,0,sa,\epsilon)$
  where $0 \leq i < n$ and $a \in \{0,1\}$. The argument is similar to
  Case 2, and requires only Formulas~\eqref{eq:character4}.
  
  \smallskip

  \noindent 
  {\bf Case 4:} $w_1$ and $w_2$ have index $(0,j+1,\epsilon,tb)$ where $0 \leq j
  < n$ and $b \in \{0,1\}$. The argument is similar to Case 2, and requires only
  Formulas~\eqref{eq:character5}.
\end{IEEEproof}

\begin{lemma}\label{lma:generateLittleMore}
  Suppose $\fA \models_{w_0} \Gamma_1 \cup \Gamma_2$.  Let $w_1, w_2$ be
  z-worlds with indices $(i_1,j_1,s_1,t_1)$ and $(i_2,j_2,s_2,t_2)$,
  respectively. Let $s^*$ be a common prefix of $s_1$ and $s_2$, and $t^*$ a
  common prefix of $t_1$ and $t_2$.  Let $i^* = \sizeof{s^*}$ and $j^* =
  \sizeof{t^*}$. Then there exists a z-world $w^*$ with index
  $(i^*,j^*,s^*,t^*)$ such that each of $w_1$ and $w_2$ is either identical to,
  or accessible from, $w^*$.
\end{lemma}

\begin{IEEEproof}
  By Lemma~\ref{lma:generate2} there exists a z-world $w^*$ with index
  $(i^*,j^*,s^*,t^*)$.  If $i^*+j^*=i_1+j_1$ then $s^*=s_1$ and $t^*=t_1$, thus
  $w^*=w_1$ by Lemma~\ref{lma:uniqueIndex}. Otherwise $i^*+j^*<i_1+j_1$ and by
  Lemma~\ref{lma:generate1}, there exists a world $w'_1$ accessible from $w^*$
  with index $(i_1,j_1,s_1,t_1)$. By Lemma~\ref{lma:uniqueIndex}, $w'_1=w_1$.
  Thus $w_1$ is accessible from $w^*$. Similarly, one can show that either
  $w^*=w_2$ or $w_2$ is accessible from $w^*$.
\end{IEEEproof}

The z-worlds of most interest are those with character $(n,n)$---of
which, by Lemmas~\ref{lma:generate2} and~\ref{lma:uniqueIndex}, there
are exactly $2^{2n}$. We refer to such worlds as \emph{g-worlds} (g for
`grid').

For any world $w$ (not just z-worlds), we define strings $s, t \in
\{0,1\}^*$ of length $n$, by setting, for all $k$ ($1 \leq k \leq n$),
$s[k] = 1$ if and only if $\fA \models_w p_k$, and $t[k] = 1$ if and
only if $\fA \models_w q_k$. We call the string $s$ the {\em
x-coordinate} of $w$, and the string $t$ its \emph{y-coordinate}.
Notice that, if $w$ is a g-world, with index $(n,n,s,t)$, then its
coordinates are $(s,t)$. The strings $s$ and $t$ may of course be
regarded as integers in the range $[0,2^n-1]$, and in the sequel we
equivocate freely between strings of length $n$ and the integers in
this range they represent.  The following abbreviations will be
useful. If $1 \leq i \leq n$, we write $p_i^*$ for $\neg p_i \wedge
p_{i+1} \wedge \cdots \wedge p_n$, and $p_i^+$ for $p_i \wedge \neg
p_{i+1} \wedge \cdots \wedge \neg p_n$.  Thus, $p_i^*$ and $p_i^+$
characterize those worlds whose $x$-coordinates are of the forms
\begin{equation}
a_1\cdots a_{i-1}0\ \overset{\text{$n-i$ times}}{\overbrace{1\cdot\cdots\cdots1}}
\quad\
a_1\cdots a_{i-1}1\ \overset{\text{$n-i$ times}}{\overbrace{0\cdot\cdots\cdots0}},
\label{eq:starPlus}
\end{equation}
respectively.  Observe that, if $s$ and $s'$ are the respective strings
(i.e.~integers) depicted in~\eqref{eq:starPlus}, then $s' = s+1$.  The
abbreviations $q_i^*$ and $q_i^+$ will be used similarly.

We now write formulas which force the g-worlds to link up into a $2^n \times
2^n$ grid (see Fig.~\ref{fig:o:worlds}). This process is complicated by the fact
that we are dealing with transitive accessibility relations.  We employ
proposition letters $o_h$, $o_v$, and refer to worlds satisfying these
proposition letters as, respectively, \emph{horizontal o-worlds} and
\emph{vertical o-worlds} (`o' stands for nothing in particular). The o-worlds'
function is to glue the g-worlds into the desired grid pattern.  Let
$\Gamma_{3,h}$ be the set of formulas:
\begin{align}
  & \Box(u_n \wedge v_n \wedge p_i^* \rightarrow \Diamond(o_h \wedge
  p_i^+)) & & (1 \leq i \leq n)
  \label{eq:standard1:h}\\
  &\Box(u_n \wedge v_n \wedge p_i^+ \rightarrow \Diamond(o_h \wedge
  p_i^+)) & & (1 \leq i \leq n)
  \label{eq:standard2:h}\\
  & \Box(u_{i-1} \wedge v_n \rightarrow \Diamond_{\leq 1}(o_h \wedge
  p_i^+)) & & (1 \leq i \leq n),
\label{eq:standard3:h} 
\end{align}
and suppose $\fA \models_{w_0} \Gamma_1 \cup \Gamma_2 \cup \Gamma_{3,h}$.
Consider a g-world $w$ with coordinates $(s,t)$. If $0 \leq s < 2^{n-1}$, then
$w$ satisfies $p_i^*$ for some $i>0$, and so has a horizontal o-world successor
by Formulas~\eqref{eq:standard1:h}; likewise, if $0 < s \leq 2^n-1$, then $w$
satisfies $p_i^+$ for some $i>0$, and so has a horizontal o-world successor by
Formulas~\eqref{eq:standard2:h}.  (Hence, if $0 < s < 2^{n-1}$, then $w$ has at
least two horizontal o-world successors.)  Finally, let $i$ be such that $1 \leq
i \leq n$, and suppose that $w^*$ is a z-world with character
$(i-1,n)$. Formulas~\eqref{eq:standard3:h} imply that there is at most one
horizontal o-world accessible from $w^*$, and satisfying $p_i^+$ (see
Fig.~\ref{fig:o:worlds:merge}).  The effect of these sets of formulas is
illustrated in Fig.~\ref{fig:o:worlds} and formalized in the following lemma:
\begin{lemma}\label{lma:standard3}
  Suppose $\fA \models_{w_0} \Gamma_1 \cup \Gamma_2 \cup \Gamma_{3,h}$.  Let $w$
  and $w'$ be g-worlds with coordinates $(s,t)$ and $(s+1,t)$,
  respectively. Then there exists a horizontal o-world $u$ accessible from both
  $w$ and $w'$ such that $\fA\models_u p_n$ if and only if $\fA\models_{w'}p_n$.
\end{lemma}

\begin{IEEEproof}
  Since $0\le s< s+1\le 2^n-1$, there exists $i$ such that $w$ satisfies
  $p_i^*$; thus $w'$ satisfies $p_i^+$. From Formulas~\eqref{eq:standard1:h}
  and~\eqref{eq:standard2:h}, there exist o-worlds $u$, $u'$ both satisfying
  $p_i^+$, with $u$ accessible from $w$, and $u'$ accessible from $w'$. Clearly,
  $\fA\models_u p_n$ if and only if $\fA\models_{w'}p_n$.  By
  Lemma~\ref{lma:generateLittleMore}, there exists a z-world $w^*$ with
  character $(i-1,n)$, for some $i$ ($1 \leq i \leq n$), such that both $w$ and
  $w'$, and hence both $u$ and $u'$, are accessible from $w^*$. From
  Formulas~\eqref{eq:standard3:h}, we have $u = u'$.
\end{IEEEproof}

Similarly, let $\Gamma_{3,v}$ be the set of formulas:
\begin{align}
  & \Box(u_n \wedge v_n \wedge q_i^* \rightarrow \Diamond(o_v \wedge
  q_i^+)) & & (1 \leq i \leq n)
  \label{eq:standard1:v}\\
  &\Box(u_n \wedge v_n \wedge q_i^+ \rightarrow \Diamond(o_v \wedge
  q_i^+)) & & (1 \leq i \leq n)
  \label{eq:standard2:v}\\
  & \Box(u_n \wedge v_{i-1}\rightarrow \Diamond_{\leq 1}(o_v \wedge
  q_i^+)) & & (1 \leq i \leq n).
\label{eq:standard3:v} 
\end{align}

\begin{lemma}\label{lma:standard4} 
  Suppose $\fA \models_{w_0} \Gamma_1 \cup \Gamma_2 \cup \Gamma_{3,v}$.  Let $w$
  and $w'$ be g-worlds with coordinates $(s,t)$ and $(s,t+1)$, respectively.
  Then there exists a vertical o-world $u$ accessible from both $w$ and $w'$
  such that $\fA\models_u q_n$ if and only if $\fA\models_{w'}q_n$.
\end{lemma}

\begin{IEEEproof}
  Analogous to Lemma~\ref{lma:standard3}.
\end{IEEEproof}

Let $\Gamma = \Gamma_1 \cup \Gamma_2 \cup \Gamma_{3,h} \cup \Gamma_{3,v}$, and
suppose $\fA \models_{w_0} \Gamma$.  Lemmas~\ref{lma:generate2}
and~\ref{lma:uniqueIndex} guarantee that, for all $s$, $t$ in the range
$[0,2^n-1]$, there exists exactly one g-world with coordinates $(s,t)$; let $G$
be the set of all these $2^{2n}$ g-worlds.  And let $O_v$, $O_h$ be sets of
horizontal and vertical o-worlds guaranteed by Lemmas~\ref{lma:standard3}
and~\ref{lma:standard4}, respectively.  Thus, the frame of $\fA$ contains, as a
subgraph, the configuration depicted in Fig.~\ref{fig:o:worlds:grid}. In short,
the formulas $\Gamma$ manufacture a $2^n \times 2^n$ grid.

Conversely, it is easy to exhibit a model of $\Gamma$, using the diagrams of
Fig.~\ref{fig:o:worlds} as our guide, containing just such a grid.

\begin{lemma}\label{lma:standard:gamma1}
  There exists a structure $\fS$ over a reflexive, transitive frame, and a world
  $w_0$ of $\fS$, such that $\fS \models_{w_0} \Gamma$.
\end{lemma}

\begin{IEEEproof}
  For $h$ and $v$ distinct symbols, define the sets:
\begin{eqnarray*}
 Z & = & 
\begin{array}[t]{@{}l@{}}
\{(i,j,s,t) \mid 0 \leq i \leq n;\ 0 \leq j \leq n;\\
\phantom{\{(i,j,s,t) \mid {}} 
s, t \in  \{0,1\}^*; \sizeof{s} = i \text{ and } \sizeof{t} = j \}
\end{array}\\
 G & = & \{(n,n, s,t) \mid s, t \in \set{0,1}^* \text{ and } \sizeof{s} = \sizeof{t} = n \}\\
 O_h\!\! & = & \{(h, s,t) \mid s, t \in \set{0,1}^*; s\notin\set{0}^*; \sizeof{s} = \sizeof{t} = n \}\\
  O_v\!\! & = & \{(v, s,t) \mid s, t \in \set{0,1}^*; t\notin\set{0}^*; \sizeof{s} = \sizeof{t} = n \}.
\end{eqnarray*} 
Note that $G \subseteq Z$.  Define the binary relations $R_Z \subseteq
Z \times Z$, $R_h \subseteq G \times O_h$ and $R_v \subseteq G \times
O_v$ by:
\begin{eqnarray*}
R_Z & = &
\begin{array}[t]{@{}l@{}}
\{\tuple{(i,j,s,t), (i',j',s',t')}\\
\qquad{}\mid i \leq i';\ j \leq j';\ s \preceq s \text{ and } t \preceq t' \}
\end{array}\\
R_h & = &
\begin{array}[t]{@{}l@{}}
\{\langle (n,n,s,t), (h,s',t') \rangle\\
\qquad{}\mid t' = t;\  s\le s'\le n \text{ and }  1\le s'\le s+1\}
\end{array}\\
R_v & = &
\begin{array}[t]{@{}l@{}}
\{\langle (n,n,s,t), (v,s',t') \rangle\\
\qquad{} \mid s' = s;\  t\le t'\le n \text{ and } 1\le t'\le t+1\}.
\end{array}
\end{eqnarray*}
Finally, let $S = Z \cup O_h \cup O_v$, and let $R_S$ be the reflexive,
transitive closure of $R_Z \cup R_h \cup R_v$.  Thus, $(S,R_S)$ is a reflexive,
transitive frame.  Define a valuation $V$ on $(S,R_S)$ by interpreting the
proposition letters as follows:
\begin{eqnarray*}
  z^\fS & = & Z;\quad
  o_h^\fS\  =\  O_h;\quad 
  o_v^\fS\  =\  O_v\\
  u_i^\fS  & = & 
  \{(i,j,s,t) \in Z \mid 0 \leq j \leq n;\ s, t \in  \{0,1\}^*\}\\
  v_j^\fS  & = & 
  \{(i,j,s,t) \in Z \mid 0 \leq i \leq n;\ s, t \in  \{0,1\}^*\}\\
  p_i^\fS  & = & 
  \{(i',j,s,t) \in Z \mid i' \geq i,\ s[i] =1\}  \cup \\
& &  \{(h,s,t) \in O_h \mid s[i] =1\} \cup\\
& &  \{(v,s,t) \in O_v \mid s[i] =1\}\\
  q_j^\fS  & = & 
  \{(i,j',s,t) \in Z \mid j' \geq j,\ t[j] =1\}  \cup \\
& &  \{(h,s,t) \in O_h \mid t[j] =1\} \cup\\
& &  \{(v,s,t) \in O_v \mid t[j] =1\}.
\end{eqnarray*}
Denote by $\fS$ the structure $(S, R_S, V)$.  Let $w_0 \in Z$ be the
element $(0,0,\epsilon,\epsilon)$. Thus, $\fS \models_{w_0} \Gamma_1$,
and, relative to $w_0$, the z-worlds of $\fS$ are simply the elements
of $Z$.  It is obvious that, for every $w = (i,j,s,t) \in Z$, the
index of $w$ is $w$ itself; moreover, for every $w = (h,s,t) \in o_h$
and every $w = (v,s,t) \in o_v$, the coordinates of $w$ are $(s,t)$.

We now show that $\fS \models_{w_0} \Gamma$.  The truth at $w_0$ of
Formulas~\eqref{eq:generate0}--\eqref{eq:standard3:v} except for
Formulas~\eqref{eq:standard3:h} and \eqref{eq:standard3:v} is
immediate.  To demonstrate the truth of
Formulas~\eqref{eq:standard3:h}, let $1 \leq i \leq n$, and fix any
world $w^*$ of $\fS$ such that $\fS \models_{w^*} u_{i-1} \wedge v_n$
(see Fig.~\ref{fig:o:worlds:merge}). We may write $w^* =
(i-1,n,s^*,t^*)$, where $\sizeof{s^*} = i-1$ and $\sizeof{t^*} =
n$. Now suppose $w'$ is any world of $\fS$ such that $\langle w^*, w'
\rangle \in R_S$ and $\fS \models_{w'} o_h \wedge p^+_i$. Again, we
may write $w' = (h,s',t')$, where $s'$ and $t'$ are bit-strings of
length $n$. We claim that $s' = s^*10 \ldots 0$ and $t' = t^*$. But
there is at most one world in $\fS$ satisfying $o_h$ and having
coordinates $(s^*10 \ldots 0, t^*)$; hence, $\fS \models_{w_0}
\Box(u_{i-1} \wedge v_n \rightarrow \Diamond_{\leq 1} (o_h \wedge
p^+_i))$, as required.

To prove the claim, observe that, by construction of $\fS$, there exists $w \in
G$ such that $\langle w^*, w \rangle \in R_S$ and $\langle w, w' \rangle \in
R_S$. Pick any such $w$ and let it have coordinates $(s,t)$. By the definition
of $R_S$ (and the fact that $\sizeof{t^*} = n$), we have: $(i)$~$t^* = t = t' $,
$(ii)$~$s^* \preceq s$, and $(iii)$~$s' = s$ or $s' = s+1$.  Referring to
Fig.~\ref{fig:o:worlds:merge}, the worlds $w^*$, $w$ and $w'$ can be reached
from $w_0$ by traversing two trees of z-worlds: an upper tree, whose leaves have
characters $(0,n)$, and a lower tree, whose elements have characters $(i,n)$
$({0 \leq i \leq n})$. The world $w^*$ in the lower tree, has character
$(i-1,n)$; $w'$ is a horizontal o-world reachable from $w^*$; $w$ is its
predecessor g-world.  Now, since $\fS \models_{w'} o_h \wedge p^+_i$, we have
$s' = s''10\ldots0$ for some string $s''$ with $\sizeof{s''} = i-1$.  Since $s$
is either $s'$ or $s'-1$, we have either $s = s''10\ldots0$ or $s =
s''01\ldots1$. Since $s^* \preceq s$ and $\sizeof{s^*}=i-1$, we have $s'' =
s^*$. Thus, $s' = s^*10 \ldots 0$ and $t' = t^*$, proving the claim.

The case of Formulas~\eqref{eq:standard3:v} is treated analogously.
\end{IEEEproof}


Now we are in a position to encode any exponential tiling problem, $(C,H,V,p)$
in our logic. We regard colours $c \in C$ as (fresh) proposition letters.
Suppose $\fA$ is transitive and $\fA \models_{w_0} \Gamma$, and let $\fA$
additionally interpret the proposition letters $c\in C$. By
Lemmas~\ref{lma:generate2}, \ref{lma:uniqueIndex}, \ref{lma:standard3},
and~\ref{lma:standard4}, the frame of $\fA$ contains the arrangement of
Fig.~\ref{fig:o:worlds:grid} as a subgraph, which we may partition into the sets
$G$ (the g-worlds), $O_h$ (the horizontal o-worlds) and $O_v$ (the vertical
o-worlds).  Intuitively, for any world $w \in G$, $c$ represents the colour of
$w$ in some (putative) tiling of $G$. Now we write formulas to ensure that the
colours 
form a tiling for $(C,H,V,p)$. Define $\Delta$ to be the following set of
formulas:
\begin{equation}
  \Box\left(u_n \wedge v_n \rightarrow \left(\bigvee C \wedge \bigwedge \{\neg
  c \vee \neg d \mid c \neq d \}\right)\right) 
  \label{eq:system0}\\
\end{equation}
\begin{align}
  &\begin{array}{@{}l@{}}
    \Box(u_n \wedge v_n \wedge \pm_1 p_n \wedge c\rightarrow\\
    \qquad\Box(o_h\land \pm_1 p_n \rightarrow c))
\end{array}
 & & (c \in C)
 \label{eq:system1}\\
  &\begin{array}{@{}l@{}}
    \Box(u_n \wedge v_n \wedge\pm_1 p_n \wedge c\rightarrow \\
    \qquad\Box(o_h\land\neg(\pm_1 p_n)\rightarrow \neg d))
\end{array}
 & & (\tuple{c,d}\notin  H)
 \label{eq:system2}\\
  &\begin{array}{@{}l@{}}
    \Box(u_n \wedge v_n \wedge \pm_1 q_n \wedge c\rightarrow\\
    \qquad\Box(o_v\land \pm_1 q_n \rightarrow c))
\end{array}
 & & (c \in C)
 \label{eq:system3}\\
  &\begin{array}{@{}l@{}}
    \Box(u_n \wedge v_n \wedge\pm_1 q_n \wedge c\rightarrow \\
    \qquad\Box(o_v\land\neg(\pm_1 q_n)\rightarrow \neg d))
\end{array}
 & & (\tuple{c,d}\notin  V).
 \label{eq:system4}
\end{align}
Formula~\eqref{eq:system0} ensures that every g-world is assigned a
unique colour. Using Lemma~\ref{lma:standard3},
Formulas~\eqref{eq:system1} ensure every horizontal o-world has the
same colour as the g-world `immediately to the right'. Together with
Formulas~\eqref{eq:system0} and~\eqref{eq:system2}, this ensures that
the g-worlds satisfy the horizontal tiling constraints. Likewise,
Formulas~\eqref{eq:system0}, \eqref{eq:system3}, and
\eqref{eq:system4} ensure that the g-worlds satisfy the vertical
tiling constraints.

\begin{lemma}\label{lma:system1}
  Suppose $\fA$ is transitive, and $\fA \models_{w_0} \Gamma \cup
  \Delta$. For all $s, t$ in the range $[0,2^n-1]$, define $f(s,t) =
  c$ if $\fA \models_w c$ for some g-world $w$ with coordinates
  $(s,t)$.  Then $f$ is well-defined, and is in fact a tiling for
  $(C,H,V)$.
\end{lemma}
\begin{IEEEproof}
Immediate.
\end{IEEEproof}

%

Now suppose $\bar{d} = d_0, \ldots, d_{m-1}$ is an $m$-tuple of elements of $C$.
Let $\pi_0$ be the formula:
\begin{equation*}
  \Box(z \wedge \neg p_1 \wedge  \cdots \wedge \neg p_n \wedge 
  \neg q_1 \wedge  \cdots \wedge \neg q_n \rightarrow d_0) 
\end{equation*}
implying that any g-world with coordinates $(0,0)$ has colour $d_0$;
and let the formulas $\pi_1$, \ldots, $\pi_{m-1}$ be defined
analogously, assigning colours $d_1$, \ldots, $d_{m-1}$ to the
g-worlds with coordinates $(1,0)$, \ldots, $(m-1,0)$. Denote by
$\Theta_{\bar{d}}$ the set of all these formulas.

\begin{lemma}\label{lma:system2}
Suppose $\fA$ is transitive, with
$\fA \models_{w_0} \Gamma \cup \Delta \cup \Theta_{\bar{d}}$, and
let the tiling $f$ be as defined in Lemma~\ref{lma:system1}.  Then
$\bar{d}$ is an initial configuration for $f$.
\end{lemma}
\begin{IEEEproof}
Immediate.
\end{IEEEproof}

Thus, we have:

\begin{lemma}\label{lma:gml:gradedK4NEXPTIMEhard}
  Let $\cK$ be any class of frames satisfying $\mbox{\rm Tr}
  \supseteq \cK \supseteq \mbox{\rm Tr} \cap \mbox{\rm Rfl}$.  The
  problem $\cGM_\cK$-{\rm Sat} is $\NEXPTIME$-hard.  It remains
  $\NEXPTIME$-hard, even when all numerical subscripts in modal
  operators are bounded by $1$.
\end{lemma}

\begin{IEEEproof}
  We reduce any exponential tiling problem $(C,H,V,p)$ to the problem
  $\cGM_\cK$-{\rm Sat}.  Fix $(C,H,V,p)$, and let an instance $\bar{d}$ of size
  $m$ be given. Write $n = p(m)$. Consider the conjunction $\phi_{\bar{d}}$ of
  all formulas in the set $\Gamma \cup \Delta \cup \Theta_{\bar{d}}$.  We claim
  that the following are equivalent: ({\em i}) $\phi_{\bar{d}}$ is satisfiable
  over $\mbox{Tr} \cap \mbox{Rfl}$; ({\em ii}) $\phi_{\bar{d}}$ is satisfiable
  over $\mbox{Tr}$; ({\em iii}) $\bar{d}$ is a positive instance of
  $(C,H,V,p)$. The implication ({\em i}) $\Rightarrow$ ({\em ii}) is
  trivial. For ({\em ii}) $\Rightarrow$ ({\em iii}), suppose $\fA \models_{w_0}
  \Gamma \cup \Delta \cup \Theta_{\bar{d}}$, with $\fA$ transitive.
  Lemmas~\ref{lma:system1} and~\ref{lma:system2} then guarantee the existence of
  a tiling $f$ of size $2^n$ for $(C,H,V)$, with initial configuration
  $\bar{d}$. For ({\em iii}) $\Rightarrow$ ({\em i}), suppose $f$ is a tiling
  for $(C,H,V)$ of size $2^n$, with initial configuration $\bar{d}$.  Taking
  $\fS$ and $w_0$ to be as in the proof of Lemma~\ref{lma:standard:gamma1}, we
  expand $\fS$ to a structure $\fS^*$ by setting $c^{\fS^*}=\set{(n,n,s,t),
    (h,s,t), (v,s,t)\mid f(s,t)=c}$ for every proposition letter $c \in C$.  It
  is obvious that $\fS^* \models_{w_0} \Delta \cup \Theta_{\bar{d}}$.
\end{IEEEproof}

Theorem~\ref{theo:main2} follows from Corollary~\ref{corollary:membership} and
Lemma~\ref{lma:gml:gradedK4NEXPTIMEhard}, noting that $\mbox{Rfl} \cap \mbox{Tr}
= \mbox{Rfl} \cap \mbox{Ser} \cap \mbox{Tr} \subseteq \mbox{Ser} \cap \mbox{Tr}
\subseteq \mbox{Tr}$.

\section{Conclusion}

In this paper, we have investigated the computational complexity of $\cGM_{\cap
  \cF}$-Sat, the satisfiability problem for graded modal logic over any frame
class $\bigcap \cF$, where $\cF \subseteq \{\mbox{Rfl}, \mbox{Ser}, \mbox{Sym},
\mbox{Tr}, \mbox{Eucl} \}$. The results are as follows. Suppose first that
$\mbox{Eucl} \not \in \cF$ and $\mbox{Tr} \not \in \cF$. Then
Theorem~\ref{theo:5:GMarbitrary} states that $\cGM_{\cap \cF}$-Sat is
$\PSPACE$-complete.  Suppose next that $\mbox{Eucl} \in \cF$ or $\{\mbox{Sym},
\mbox{Tr}\} \subseteq \cF$.  Then Theorem~\ref{theo:main1} states that
$\cGM_{\cap \cF}$-Sat is $\NP$-complete.  Suppose finally that $\mbox{Eucl},
\mbox{Sym} \not \in \cF$, but $\mbox{Tr} \in \cF$. Then Theorem~\ref{theo:main2}
states that $\cGM_{\cap \cF}$-Sat is $\NEXPTIME$-complete. All these results
hold under both unary and binary coding of numerical subscripts.

\section*{Acknowledgment}
The authors would like to thank Prof.~Ulrike~Sattler for helpful
discussions. Yevgeny Kazakov is supported by the EPSRC, grant number EP/G02085X.
Ian Pratt-Hartmann is supported by the EPSRC, grant number EP/F069154.



%
\bibliographystyle{IEEETran}
\bibliography{gml-tr}
\end{document}